\documentclass[12pt,preprint]{aastex}

\usepackage{amssymb}
\usepackage{amsmath}
\usepackage{graphicx}
\usepackage{subfigure}
\usepackage{latexsym}
\usepackage{xcolor}
\usepackage{float}

\shorttitle{Kelvin-Helmholtz instability in rotating solar jets}
\shortauthors{Zaqarashvili et al.}

\begin{document}

\title{Stability of rotating magnetized jets in the solar atmosphere. \\ I. Kelvin-Helmholtz instability}

\author{Teimuraz V.~Zaqarashvili\altaffilmark{1,4,5}, Ivan Zhelyazkov\altaffilmark{2}, and Leon Ofman\altaffilmark{3,6,7}}

\altaffiltext{1}{Institute of Physics, IGAM, University of Graz, Universit\"atsplatz 5, 8010, Graz, Austria \\
\hspace*{6.3mm}Email: teimuraz.zaqarashvili@uni-graz.at}
\altaffiltext{2}{Faculty of Physics, Sofia University, 5 James Bourchier Blvd., 1164 Sofia, Bulgaria}
\altaffiltext{3}{Catholic University of America, Washington, DC 20064, USA}
\altaffiltext{4}{Abastumani Astrophysical Observatory at Ilia State University, 0162 Tbilisi, Georgia}
\altaffiltext{5}{Space Research Institute, Austrian Academy of Sciences, Schmiedlstrasse 6, 8042 Graz, Austria}
\altaffiltext{6}{NASA Goddard Space Flight Center, Greenbelt, MD 20771, USA}
\altaffiltext{7}{Visiting, Department of Geosciences, Tel Aviv University, Tel Aviv 69978, Israel}

\begin{abstract}

Observations show various jets in the solar atmosphere with significant rotational motions, which may undergo instabilities leading to heat ambient plasma. We study the Kelvin-Helmholtz (KH) instability of twisted and rotating jets caused by the velocity jumps near the jet surface. We derive a dispersion equation with appropriate boundary condition for total pressure (including centrifugal force of tube rotation), which governs the dynamics of incompressible jets. Then, we obtain analytical instability criteria of Kelvin-Helmholtz instability in various cases, which were verified by numerical solutions to the dispersion equation. We find that twisted and rotating jets are unstable to KH instability when the kinetic energy of rotation is more than the magnetic energy of the twist. Our analysis shows that the azimuthal magnetic field of $1$--$5$~G can stabilize observed rotations in spicule/macrospicules and X-ray/EUV jets. On the other hand, non-twisted jets are always unstable to KH instability. In this case, the instability growth time is several seconds for spicule/macrospicules and few minutes (or less) for EUV/X-ray jets. We also find that standing kink and torsional Alfv\'en waves are always unstable near the antinodes due to the jump of azimuthal velocity at the surface, while the propagating waves are generally stable. KH vortices may lead to enhanced turbulence development and heating of surrounding plasma, therefore rotating jets may provide energy for chromospheric and coronal heating.

\end{abstract}


\section{Introduction}\label{intro}

Jets are ubiquitous in the solar atmosphere. Observations show various kinds of motions with different temperature, speed and other plasma parameters such as: spicules, X-ray jets, EUV (extreme-ultraviolet) jets, chromospheric anemone jets, tornadoes, etc. Spicules appear as cool dense chromospheric plasma jets rising up into the solar corona with mean speed of $20$--$25$~km\,s$^{-1}$ and they are usually detected in chromospheric H$\alpha$, D$_3$ and Ca {\footnotesize\textsc{II}} H lines \citep{Beckers1968}. Recent Hinode observations, however, revealed another type of spicules (Type II spicules) with much higher ($50$--$100$~km\,s$^{-1}$) speeds \citep{De Pontieu2007}. X-ray jets were discovered by the YOHKOH Soft X-ray Telescope \citep{Shibata1992,Shimojo1996} and have been observed in details by the Hinode X-Ray Telescope \citep{Savcheva2007}. They are hot jets with speeds of around $200$--$600$~km\,s$^{-1}$, which are mainly associated with small C-class flares. Extreme-ultraviolet (EUV) jets look like X-ray jets, but are smaller and shorter lived structures \citep{Chae1999}. Chromospheric anemone jets were discovered by Solar Optical Telescope (SOT) onboard the Hinode spacecraft \citep{Shibata2007, Nishizuka2011}. They are typically $2$--$5$~Mm long with the mean speed of $10$--$20$~km\,s$^{-1}$ and they are probably connected to small scale magnetic reconnection in the solar chromosphere. Recent high resolution observations from ROSA (Rapid Oscillations in the Solar Atmosphere) multi wavelength imager on the Dunn Solar Telescope showed numerous small-scale H$\alpha$ jets, which are associated with small scale explosive events in the chromosphere \citep{Kuridze2011}.

Many of the observed jets in the solar atmosphere show rotational motions. Early spectral observations showed a tilt of spicule spectra, which has been explained by the rotation of spicules around their axes \citep{Pasachoff1968,Rompolt1975}. Type II spicules also show significant torsional motions \citep{De Pontieu2012,Tian2014}. SOHO Coronal Diagnostic Spectrometer (CDS) showed rotating transition region features called macrospicules \citep{Pike1998}. Follow up observations confirmed that macrospicules are really rotating structures \citep{Kamio2010,Curdt2012}. \citet{Moore2013} performed systematic analysis of 54 polar X-ray jets, which showed significant rotational motion \citep{Liu2011,Cheung2015}. Atmospheric Imaging Assembly (AIA) aboard the Solar Dynamics Observatory (SDO) revealed swirling EUV jets \citep{Zhang2014}. Recent Hinode observations also show the rotational motion of chromospheric jets \citep{Liu2009}. Therefore, some of observed jets in the solar chromosphere and corona exhibit significant rotational motion around their axes. Recent numerical simulations also show the rotational signature in modeled jets in various situations \citep{Kitiashvili2013,archontis2013,Fang2014,Pariat2015,Lee2015}.

Tornado-like structures have been reported in the solar atmosphere since 19th century \citep{Zolner1869}. They have been associated with filaments/prominences, which are called tornado prominences \citep{Pettit1950}. Large vortex flows at the photosphere \citep{Brandt1988,Attie2009} or the expansion of a twisted flux rope into the coronal cavity \citep{li2012,Panesar2013} were assumed to be responsible for the formation of solar giant tornadoes, that may play a distinct role in the supply of mass and twists to filaments. On the other hand, the appearances of smaller scale tornadoes (magnetic tornadoes) in the chromosphere were recently observed \citep{Wedemeyer2009,wedemeyer2012}, which may provide a mechanism for channeling energy from the lower into the upper solar atmosphere. The small scale tornadoes are probably formed by small scale photospheric vortex flows, which are reported in observations \citep{Bonet2008,Bonet2010,Steiner} and numerical simulations \citep{Stein,Shelyag2011}.

Axial motion and/or rotation of magnetic flux tubes lead to the velocity jump at the tube boundary, which may eventually results in the Kelvin-Helmholtz instability (KHI) under some circumstances. Kelvin-Helmholtz (KH) vortices have been intensively observed in recent years in solar prominences \citep{Berger2010,Ryutova2010} and at boundaries of rising coronal mass ejections (CMEs) \citep{Ofman2010,Ofman2011,Foullon2011,Foullon2013,Mostl2013,Nykyri2013} as well as in streamers \citep{Feng2013}. The observations resulted in increased interest toward KHI in magnetic flux tubes. The KHI has been studied in twisted magnetic flux tubes moving in nonmagnetic environment \citep{zaqarashvili2010}, magnetic tubes with partially ionized plasmas \citep{Soler2012,Martinez2015}, in spicules \citep{Zaqarashvili2011,Zhelyazkov2012a}, in soft X-ray jets \citep{Zhelyazkov2012b}, in photospheric tubes \citep{Zhelyazkov2012}, in high-temperature and cool surges \citep{Zhelyazkov2015a,Zhelyazkov2015b}, and at the boundary of rising CMEs \citep{Mostl2013, Zhelyazkov2015c}. Azimuthal velocity component in magnetohydrodynamic (MHD) kink waves may also lead to KHI in coronal loops \citep{Ofman1994,Terradas2008,Soler2010,Antolin2014}.

KH vortices can be considered as one of the important sources for MHD turbulence in the solar wind (e.g., \citealp{Bavassano1989}). The KHI can be developed by velocity discontinuity at boundaries of magnetic flux tubes owing to the relative motion of the tubes with regards to the solar wind or neighboring tubes.  However, the flow-aligned magnetic field may suppress KHI for typical velocity jumps at boundaries of observed magnetic structures, which is generally sub-Alfv\'enic. The KHI can still survive in the twisted tubes in nonmagnetic or twisted environment as the harmonics with sufficiently large azimuthal wave mode number $m$ are unstable for any sub-Alfv\'enic flows along the tubes \citep{zaqarashvili2010,zaqarashvili2014a}. The transverse motion of non-twisted magnetic tubes may also result in the KHI for any value of speed, which may lead to the plasma turbulence and heating. Recently observed fast disappearance of rapid redshifted and blueshifted excursions (RREs, RBEs) in the H$\alpha$ line has been interpreted by the heating of the structures due to the KHI at their boundary during transverse motion \citep{Kuridze2015}.

In general, any velocity component across the magnetic fields inside and outside a flux tube may lead to the fast development of KHI. Therefore, the rotation of non-twisted magnetic tubes is generally unstable to KHI \citep{Bodo1989,Bodo1996,Terradas2008,Soler2010}. The twist of tube magnetic field may, however, stabilize the instability. \citet{Soler2010} made the first attempt to consider the stabilizing effect of twist on KHI assuming a slab model. They concluded that a very weak twist of magnetic
field lines could be enough to prevent the triggering of the KHI in a cylindrical coronal loop. On the other hand, the consideration of cylindrical symmetry could be important as boundary conditions at a tube surface (namely Lagrangian total pressure) contain the contribution from the tension force of magnetic field azimuthal component \citep{dungey1954,bennett1999,zaqarashvili2010,zaqarashvili2014a}. Equally important, the contribution from the centrifugal force of rotation should be taken into account in the boundary condition of Lagrangian total pressure, which was neglected in previous studies.

Here we study the KHI of twisted and rotating magnetic jets in the approximation of incompressible magnetohydrodynamics (MHD) using the appropriate boundary condition of Lagrangian total pressure. Obtained results have direct applications to the stability of various jets, spicules and tornadoes observed in the solar atmospheres.  The paper is organized as follows. The dispersion equation governing the dynamics of twisted and rotating jets is obtained in Sect.~2. Analytical instability criteria for the KHI are derived in Sect.~3. Section 4 deals with the numerical solutions to the wave dispersion equation. Connection of obtained KHI criteria to various jets observed in the solar atmosphere is discussed in Sect.~5. The main results are summarized in the Sect.~6.

\section{Main equations}

We consider a magnetized helical jet of radius $a$ in the cylindrical coordinate system $(r,\theta,z)$, where $r$ is the distance from the jet axis. The magnetic and velocity fields inside the jet are supposed to be
\begin{equation}
\label{magnetic}(0,B_{\theta}(r),B_{z{\rm i}})
\end{equation}
and
\begin{equation}
\label{velocity}(0,U_{\theta}(r),U_z),
\end{equation}
respectively. This configuration also describes twisted and rotating magnetic tubes moving along their axes, therefore we use both terms (tube and jet) in the remaining part of the paper. For simplicity we suppose that the jet has homogeneous density, $\rho_\mathrm{i}=\mathrm{ const}$, uniform axial component of the magnetic field, $B_{z{\rm i}}$, and uniform axial velocity, $U_z$. The jet is surrounded by homogeneous medium with $\rho_\mathrm{e}=\mathrm{const}$, uniform axial magnetic field, $(0,0,B_\mathrm{e})$ and zero velocity field. Pressure balance condition inside the jet yields the following radial profile of the total pressure
\begin{equation}
\label{pressure}
p_{\rm t}(r)=p(r)+ {{B^2_{z{\rm i}}}\over {8 \pi}} + {{B^2_{\theta}(r)}\over {8 \pi}}=p_{\rm t}(0)-{{1}\over {4 \pi}}\int{{{B^2_{\theta}(r)}\over {r}}\mathrm{d}r}+ \int{{{\rho_\mathrm{i}U^2_{\theta}(r)}\over {r}}\mathrm{d}r},
\end{equation}
where $p_{\rm t}$ is the total (thermal $+$ magnetic) pressure and $p$ is the thermal pressure.

Linearized MHD equations, which govern the incompressible dynamics of perturbations are
\begin{equation}
\label{1}
\left [{{\partial }\over {\partial t}}+{{U_{\theta}}\over {r}}{{\partial }\over {\partial \theta}}+ U_z {{\partial }\over {\partial z}}\right ]u_r-2{{U_{\theta}}\over {r}} u_{\theta}-
{{1}\over {4 \pi \rho_\mathrm{i}}}\left [{{B_{\theta}}\over {r}}{{\partial }\over {\partial \theta}}+ B_{z{\rm i}}{{\partial }\over {\partial z}}\right ]b_r+ 2{{B_{\theta}}\over {4 \pi \rho_\mathrm{i}r}}b_{\theta}=-{{1}\over {\rho_\mathrm{i}}}{{\partial p_{\rm t}}\over {\partial r}},
\end{equation}
\begin{eqnarray}
\label{2}
\left [{{\partial }\over {\partial t}}+{{U_{\theta}}\over {r}} {{\partial }\over {\partial \theta}}+ U_z {{\partial }\over {\partial z}}\right ]u_{\theta}+{{1}\over {r}}{{\partial (rU_{\theta})}\over {\partial r}}u_{r}-
{{1}\over {4 \pi \rho_\mathrm{i}}}\left [{{B_{\theta}}\over {r}}{{\partial }\over {\partial \theta}}+ B_{z{\rm i}}{{\partial }\over {\partial z}}\right ]b_{\theta}-{{1}\over {4 \pi \rho_\mathrm{i}}}{{1}\over {r}}{{\partial (rB_{\theta})}\over {\partial r}}b_r \nonumber \\
  \nonumber \\
{}= -{{1}\over {\rho_\mathrm{i}r}}{{\partial p_{\rm t}}\over {\partial \theta}},
\end{eqnarray}
\begin{equation}
\label{3}
\left [{{\partial }\over {\partial t}}+{{U_{\theta}}\over {r}} {{\partial }\over {\partial \theta}}+ U_z {{\partial }\over {\partial z}}\right ]u_z-
{{1}\over {4 \pi \rho_\mathrm{i}}}\left [{{B_{\theta}}\over {r}}{{\partial }\over {\partial \theta}}+ B_{z{\rm i}}{{\partial }\over {\partial z}}\right ]b_z=-{{1}\over {\rho_\mathrm{i}}}{{\partial p_{\rm t}}\over {\partial z}},
\end{equation}
\begin{equation}
\label{4}
\left [{{\partial }\over {\partial t}}+{{U_{\theta}}\over {r}} {{\partial }\over {\partial \theta}}+ U_z {{\partial }\over {\partial z}}\right ]b_r-
\left [{{B_{\theta}}\over {r}}{{\partial }\over {\partial \theta}}+ B_{z{\rm i}}{{\partial }\over {\partial z}}\right ]u_r=0,
\end{equation}
\begin{equation}
\label{5}
\left [{{\partial }\over {\partial t}}+{{U_{\theta}}\over {r}} {{\partial }\over {\partial \theta}}+ U_z {{\partial }\over {\partial z}}\right ]b_{\theta}-r{{\partial }\over {\partial r}}\left ({{U_{\theta}}\over {r}}\right )b_r-\left [{{B_{\theta}}\over {r}}{{\partial }\over {\partial \theta}}+ B_{z{\rm i}}{{\partial }\over {\partial z}}\right ]u_{\theta}+r{{\partial }\over {\partial r}}\left ({{B_{\theta}}\over {r}}\right )u_r=0,
\end{equation}
\begin{equation}
\label{6}
\left [{{\partial }\over {\partial t}}+{{U_{\theta}}\over {r}} {{\partial }\over {\partial \theta}}+ U_z {{\partial }\over {\partial z}}\right ]b_z-
\left [{{B_{\theta}}\over {r}}{{\partial }\over {\partial \theta}}+ B_{z{\rm i}}{{\partial }\over {\partial z}}\right ]u_z=0,
\end{equation}
where $b_r,b_{\theta},b_z$ are the perturbations of magnetic field, $u_r,u_{\theta},u_z$ are the perturbations of velocity and $p_{\rm t}$ is the perturbation of total pressure. In addition, we have solenoidal equations for velocity and magnetic field perturbations
\begin{equation}
\label{7}
{{\partial u_r}\over {\partial r}}+ {{u_r}\over {r}}+{{1}\over {r}}{{\partial u_{\theta}}\over {\partial \theta}}+ {{\partial u_z}\over {\partial z}}=0,
\end{equation}
\begin{equation}
\label{8}
{{\partial b_r}\over {\partial r}}+ {{b_r}\over {r}}+{{1}\over {r}}{{\partial b_{\theta}}\over {\partial \theta}}+ {{\partial b_z}\over {\partial z}}=0.
\end{equation}

We search the solutions to the equations in normal modes in the form
\begin{equation}
\label{Fourier}
\exp[\mathrm{i}(-\omega t + m\theta + k_z z)],
\end{equation}
where $\omega$ is the frequency (in general, can be complex), $m$ is the azimuthal wave mode number (an integer, that can be positive, negative or zero) and $k_z$ is the longitudinal wavenumber.

The substitution of Eq.~(\ref{Fourier}) into Eqs.~(\ref{1})--(\ref{8}) yields the following equations:
\begin{equation}
\label{pert1}
-\mathrm{i}\sigma u_r -2{{U_{\theta}}\over {r}} u_{\theta}-{{\mathrm{i} \mu}\over {4 \pi \rho_\mathrm{i}}}b_r+2{{B_{\theta}}\over {4 \pi \rho_\mathrm{i} r}}b_{\theta}=-{{1}\over {\rho_\mathrm{i}}}{{\mathrm{d} p_{\rm t}}\over {\mathrm{d}r}},
\end{equation}
\begin{equation}
\label{pert2}
-\mathrm{i}\sigma u_{\theta}+{{1}\over {r}}{{\mathrm{d}(rU_{\theta})}\over {\mathrm{d}r}}u_{r}-
{{\mathrm{i} \mu}\over {4 \pi \rho_\mathrm{i}}}b_{\theta}-{{1}\over {4 \pi \rho_\mathrm{i}}}{{1}\over {r}}{{\mathrm{d}(rB_{\theta})}\over {\mathrm{d}r}}b_r=
-{{\mathrm{i}m}\over {\rho_\mathrm{i}r}}p_{\rm t},
\end{equation}
\begin{equation}
\label{pert3}
-\mathrm{i}\sigma u_z-{{\mathrm{i} \mu}\over {4 \pi \rho_\mathrm{i}}}b_z=-{{\mathrm{i}k_z}\over {\rho_\mathrm{i}}}p_{\rm t},
\end{equation}
\begin{equation}
\label{pert4}
-\mathrm{i}\sigma b_r - \mathrm{i} \mu u_r=0,
\end{equation}
\begin{equation}
\label{pert5}
-\mathrm{i}\sigma b_{\theta}-r{{\mathrm{d}}\over {\mathrm{d}r}}\left ({{U_{\theta}}\over {r}}\right )b_r-\mathrm{i} \mu u_{\theta}+r{{\mathrm{d}}\over {\mathrm{d}r}}\left ({{B_{\theta}}\over {r}}\right )u_r=0,
\end{equation}
\begin{equation}
\label{pert6}
-\mathrm{i}\sigma b_z-\mathrm{i} \mu u_z=0,
\end{equation}
\begin{equation}
\label{pert7}
{{\mathrm{d}u_r}\over {\mathrm{d}r}}+ {{u_r}\over {r}}+{{\mathrm{i}m}\over {r}}u_{\theta}+ \mathrm{i}k_z u_z=0,
\end{equation}
where
\begin{equation}
\label{sigma}
\sigma=\omega-m{{U_{\theta}}\over {r}}-k_zU_z
\end{equation}
is the Doppler-shifted frequency and
\begin{equation}
\label{sigma1}
\mu=m{{B_{\theta}}\over {r}}+k_zB_{z{\rm i}}.
\end{equation}

It is now convenient to use the Lagrangian displacement, $\boldsymbol \xi$, which is defined as follows \citep{chandrasekhar1961}
\begin{equation}
\label{xi}
u_r=-\mathrm{i}\sigma \xi_r, \quad u_{\theta}=-\mathrm{i}\sigma \xi_{\theta} - r{{\mathrm{d}}\over {\mathrm{d}r}}\left ({{U_{\theta}}\over {r}}\right )\xi_r, \,\, u_z= -\mathrm{i}\sigma \xi_z.
\end{equation}

In terms of $\boldsymbol \xi$, Eqs.~(\ref{pert1})--(\ref{pert7}) are rewritten as
\begin{equation}
\label{xi1}
\left [\sigma^2-\omega^2_\mathrm{Ai}-r{{\mathrm{d}}\over {\mathrm{d}r}}\left ({{U^2_{\theta}}\over {r^2}}\right )+{{r}\over {4 \pi \rho_\mathrm{i}}}{{\mathrm{d}}\over {\mathrm{d}r}}\left ({{B^2_{\theta}}\over {r^2}}\right ) \right ]\xi_r-2i\left [\sigma {{U_{\theta}}\over {r}}+ {{B_{\theta}\mu}\over {4 \pi \rho_\mathrm{i} r}}\right ]\xi_{\theta}={{1}\over {\rho_\mathrm{i}}}{{\mathrm{d} p_{\rm t}}\over {\mathrm{d}r}},
\end{equation}
\begin{equation}
\label{xi2}
\left [\sigma^2-\omega^2_\mathrm{Ai}\right ]\xi_{\theta}+2i\left [\sigma {{U_{\theta}}\over {r}}+ {{B_{\theta}\mu}\over {4 \pi \rho_\mathrm{i} r}}\right ]\xi_{r}={{\mathrm{i}m}\over {\rho_\mathrm{i}r}}p_{\rm t},
\end{equation}
\begin{equation}
\label{xi3}
\left [\sigma^2-\omega^2_\mathrm{Ai}\right ]\xi_{z}={{\mathrm{i}k_z}\over {\rho_\mathrm{i}}}p_{\rm t},
\end{equation}
\begin{equation}
\label{x4}
{{\mathrm{d}\xi_r}\over {d\mathrm{d}}}+ {{\xi_r}\over {r}}+{{\mathrm{i}m}\over {r}}\xi_{\theta}+ \mathrm{i}k_z \xi_z=0,
\end{equation}
where
\begin{equation}
\label{omega_A}
\omega_\mathrm{Ai}={{\mu}\over {\sqrt{4 \pi \rho_\mathrm{i}}}}
\end{equation}
is the local Alfv\'en frequency inside the jet.

Excluding $\xi_{\theta}$ and $\xi_{z}$ from these equations we obtain
\begin{equation}
\label{two1}
\rho_\mathrm{i}\left (\sigma^2-\omega^2_\mathrm{Ai}\right )\left [{{\mathrm{d}\xi_r}\over {\mathrm{d}r}}+ {{\xi_r}\over {r}}\right ]+{{2m\rho_\mathrm{i}d_2}\over {r}}\xi_{r}=\left ({{m^2}\over {r^2}}+k_z^2\right )p_{\rm t},
\end{equation}
\begin{equation}
\label{two2}
\rho_\mathrm{i}d_1\xi_r=\left (\sigma^2-\omega^2_\mathrm{Ai}\right ){{\mathrm{d} p_{\rm t}}\over {\mathrm{d}r}}-2md_2{{p_{\rm t}}\over r},
\end{equation}
where
\begin{eqnarray}
\label{d2}
d_1=\left (\sigma^2-\omega^2_\mathrm{Ai}\right )^2-r\left (\sigma^2-\omega^2_\mathrm{Ai}\right )\left [{{\mathrm{d}}\over {\mathrm{d}r}}\left ({{U^2_{\theta}}\over {r^2}}\right )-{{1}\over {4 \pi \rho_\mathrm{i}}}{{\mathrm{d}}\over {\mathrm{d}r}}\left ({{B^2_{\theta}}\over {r^2}}\right )\right ] -4d^2_2 , \nonumber \\
  \nonumber \\
d_2=\sigma {{U_{\theta}}\over {r}}+ {{B_{\theta}\mu}\over {4 \pi \rho_\mathrm{i} r}}.
\end{eqnarray}

These two equations can be cast into one equation
\begin{equation}
\label{gen}
\left [\left (\sigma^2-\omega^2_\mathrm{Ai}\right ){{\mathrm{d}}\over {\mathrm{d}r}} +{{\sigma^2-\omega^2_\mathrm{Ai}+2md_2}\over {r}} \right ]\left [{{\sigma^2-\omega^2_\mathrm{Ai}}\over {d_1}}{{\mathrm{d} p_{\rm t}}\over {\mathrm{d}r}}-{{2md_2}\over {d_1}}{{p_{\rm t}}\over {r}} \right ]-\left ({{m^2}\over {r^2}}+k_z^2\right )p_{\rm t}=0.
\end{equation}

Now we consider that the rotation and the magnetic twist of jet are homogeneous, that is,
\begin{equation}
\label{hom}
U_{\theta}=r\Omega, \,\,\, B_{\theta}=rA,
\end{equation}
where $\Omega$ and $A$ are constants. These profiles significantly simplify Eq.~(\ref{gen}), which is now readily transformed into the Bessel equation
\begin{equation}
\label{bessel}
{{\mathrm{d}^2 p_{\rm t}}\over {d\mathrm{d}^2}}+{{1}\over {r}}{{\mathrm{d} p_{\rm t}}\over {\mathrm{d}r}}-\left ({{m^2}\over {r^2}}+m^2_\mathrm{i}\right )p_{\rm t}=0,
\end{equation}
where
\begin{equation}
\label{bessel1}
m^2_\mathrm{i}=k_z^2\left [1-4\left ({{\sigma \Omega+{{A\omega_\mathrm{Ai}}/{\sqrt{4 \pi \rho_\mathrm{i}}}}}\over {\sigma^2-\omega^2_\mathrm{Ai}}}\right )^2\right ].
\end{equation}

The equation governing the dynamics of plasma outside the jet (without twist and velocity field, that is, $A=0$, $\Omega=0$ and $U_z=0$) is the same Bessel equation but  $m_\mathrm{i}$ is replaced by $k_z$.

Inside the jet, the solution to Eq.~(\ref{bessel}) bounded on the jet axis is the modified Bessel function of first kind
\begin{equation}
\label{inside}
p_{\rm t}=a_{\rm i} I_m(m_\mathrm{i}r),
\end{equation}
where $a_{\rm i}$ is a constant.

Outside the jet, the solution bounded in infinity is the modified Bessel function of the second kind
\begin{equation}
\label{outside}
p_{\rm t}=a_{\rm e} K_m(k_zr),
\end{equation}
where $a_{\rm e}$ is a constant.

To get the dispersion equation governing the dynamics of the jet one needs to merge the solutions at the jet surface through boundary conditions. Boundary conditions of perturbations in rotating and twisted tubes need special attention; therefore we dedicate a separate subsection to this problem.

\subsection{Boundary conditions in rotating and twisted tubes}

Non-twisted and non-rotating tubes yield the continuity of Lagrangian radial displacement and total pressure at the tube surface i.e.
\begin{equation}
\label{boundary1}
[\xi_r]_a=0,\,\,[p_{\rm t}]_a=0.
\end{equation}
When the tube is twisted and the twist has discontinuity at the tube surface then the Lagrangian total pressure is continuous and the boundary conditions are \citep{dungey1954,bennett1999,zaqarashvili2010,zaqarashvili2014a}
\begin{equation}
\label{boundary2}
[\xi_r]_a=0,\,\,\left [p_{\rm t} -{{B^2_{\theta}}\over {4 \pi a}}\xi_r \right ]_a = 0.
\end{equation}
It is readily followed from Eq.~(\ref{two2}) for $U_{\theta}=0$. Multiplying that equation by $\mathrm{d}r$ and considering the limit of $\mathrm{d}r \rightarrow 0$ through boundary $r=a$, one can find the relation $d(p_{\rm t}-(B^2_{\theta}/4\pi a)\xi_r)=0$, which gives the second boundary condition in Eq.~(\ref{boundary2}). The second term in the boundary condition for the Lagrangian total pressure stands for the pressure from the magnetic tension force.

If non-twisted ($B_{\theta}=0$) tube rotates and the rotation has discontinuity at the tube surface then the same procedure leads to the boundary condition for the Lagrangian total pressure
\begin{equation}
\label{boundary3}
\left [p_{\rm t} + {{\rho_\mathrm{i}U^2_{\theta}}\over a} \xi_r \right ]_a = 0.
\end{equation}
Here the second term describes the contribution of centrifugal force in the pressure balance, which has been usually neglected in the boundary condition of total pressure in rotating tubes \citep{Bodo1989,Bodo1996,Terradas2008,Soler2010}. However, this term appears in the instability criterion of KHI, therefore one should keep it in the following calculations.

If a magnetic tube is twisted and rotating then Eq.~(\ref{two2}) gives the boundary condition for the Lagrangian total pressure as
\begin{equation}
\label{boundary4}
\left [p_t +\left ({{\rho_\mathrm{i}U^2_{\theta}}\over a}-{{B^2_{\theta}}\over {4 \pi a}}\right )\xi_r \right ]_a=0.
\end{equation}
In the case of homogeneous rotation and twist (Eq.~\ref{hom}) the correct expressions of continuity of Lagrangian radial displacement and total pressure at the tube surface are
\begin{equation}
\label{boundary5}
[\xi_r]_a=0,\,\,\left [p_t +a\left (\rho_\mathrm{i}\Omega^2-{{A^2}\over {4 \pi}}\right )\xi_r \right ]_a=0.
\end{equation}
We will use these expressions to obtain a dispersion equation governing the incompressible perturbations of helical magnetic jets.

\subsection{Dispersion equation}

Using the boundary conditions Eq.~(\ref{boundary5}) we obtain the dispersion equation
\begin{eqnarray}
\label{dispersion}
{{(\sigma^2-\omega^2_\mathrm{Ai})F_m(m_\mathrm{i}a)-2m(\sigma \Omega+A\omega_\mathrm{Ai}/\!\sqrt{4 \pi \rho_\mathrm{i}})}\over {\rho_\mathrm{i}(\sigma^2
-\omega^2_\mathrm{Ai})^2-4\rho_\mathrm{i}(\sigma \Omega+A\omega_\mathrm{Ai}/\!\sqrt{4 \pi \rho_\mathrm{i}})^2}} \nonumber \\
  \nonumber \\
{}={{P_{m}(k_z a)}\over {\rho_{\mathrm{e}}(\omega^2-\omega^2_\mathrm{Ae})-(\rho_\mathrm{i}\Omega^2-A^2/4 \pi )P_{m}(k_z a)}},
\end{eqnarray}
where
\[
F_{m}(m_\mathrm{i}a)={{m_\mathrm{i}aI^{\prime}_{m}(m_\mathrm{i}a)}\over {I_{m}(m_\mathrm{i}a)}},\qquad P_{m}(k_z a)={{k_z aK^{\prime}_{m}(k_z a)}\over {K_{m}(k_z a)}},\qquad \omega_\mathrm{Ae}={{k_zB_{\rm e}}\over {\sqrt{4 \pi \rho_\mathrm{e}}}}.
\]

This dispersion equation governs the dynamics of incompressible normal modes in twisted and rotating magnetic jets. For the zero velocity field, $\Omega=0, U_z=0$, one may recover the dispersion equation governing static twisted tubes (Eq.~(23) in \citealp{bennett1999}). For zero rotation, $\Omega=0$, one may recover
the dispersion equation governing the twisted tubes with axial flows (Eq.~(25) in \citealp{zaqarashvili2014a}).

In general, the frequency in Eq.~(\ref{dispersion}) is a complex quantity, $\mathrm{Re}(\omega) + \mathrm{i}\,\mathrm{Im}(\omega)$, where real and imaginary parts correspond to the frequency and growth rate of unstable harmonics, respectively. Two types of instability may develop in rotating twisted tubes: kink instability and KHI. The kink instability arises in twisted tubes when the twist exceeds a threshold value. KHI instability arises at the interface of two fluid layers which move with different speeds. Flows are generally uniform in both layers, but strong velocity shear arises near the interface. The velocity shear forms a vortex sheet, which consequently becomes unstable to spiral-like perturbations. When a magnetic tube moves along or rotates around its axis, then a vortex sheet is formed near the tube boundary in both cases, which may become unstable to KHI under certain conditions. Normal mode analysis \citep{dungey1954} and energy consideration method \citep{lundquist1951} show the similar thresholds of the kink instability in twisted magnetic tubes as $B_{\theta}(a)>2B_{z{\rm i}}$.  This leads to the critical twist angle of ${\sim}65^\circ$.  External magnetic field may increase the threshold and thus stabilize the instability \citep{bennett1999}.  On the other hand, a flow along the twisted magnetic tube may decrease the threshold \citep{zaqarashvili2010,zaqarashvili2014b}. But, above-mentioned criterion is quite good for twisted tubes in general. Therefore, the tubes with weaker twist are stable for the kink instability. In the following, we consider only weakly twisted tubes, whose twist is well below the instability threshold, therefore only KHI remains in the system.

The dispersion equation, Eq.~(\ref{dispersion}), can be solved analytically and numerically. First, we solve the equation analytically in thin tube (or long wavelength) approximation. Then we solve the equation numerically and compare the obtained results.

\section{Analytical instability criteria for Kelvin-Helmholtz instability}

The long wavelength approximation, $k_z a \ll 1$, yields $m_\mathrm{i}a \ll 1$, therefore we have
\[
F_m(m_\mathrm{i}a)=\frac{m_\mathrm{i} aI^{\prime}_m(m_\mathrm{i}a)}{I_m(m_\mathrm{i}a)}\approx |m|
\]
and
\[
P_m(ka) = \frac{k_z aK^{\prime}_m(k_z a)}{K_m(k_z a)}\approx -|m|.
\]
Then Eq.~(\ref{dispersion}) is reduced to the polynomial dispersion relation
\begin{eqnarray}
\label{disp1}
\left [\sigma^2-\omega^2_\mathrm{Ai}-2\left (\sigma \Omega+A\omega_\mathrm{Ai}/\!\sqrt{4 \pi \rho_\mathrm{i}}\right ){{m}\over {|m|}}\right ]\left[\rho_\mathrm{e}\left (\omega^2-\omega^2_\mathrm{Ae}\right )+\left (\rho_\mathrm{i}\Omega^2-A^2/4 \pi \right )|m|\right ] \nonumber \\
  \nonumber \\
{}= -\rho_\mathrm{i}
\left [(\sigma^2-\omega^2_\mathrm{Ai})^2-4(\sigma \Omega+A\omega_\mathrm{Ai}/\!\sqrt{4 \pi \rho_\mathrm{i}})^2\right ].
\end{eqnarray}
This equation can be significantly simplified for positive or negative azimuthal wave mode number, $m$. We consider these two cases separately.

For a positive azimuthal wave mode number, $m>0$, both sides of Eq.~(\ref{disp1}) can be divided by $\sigma^2-\omega^2_\mathrm{Ai}-2(\sigma \Omega+A\omega_\mathrm{Ai}/\!\sqrt{4 \pi \rho_\mathrm{i}})\neq0$ and we obtain a second order polynomial
\begin{equation}\label{disp2}
\left (1+{{\rho_\mathrm{e}}\over {\rho_\mathrm{i}}} \right )\omega^2 - 2(k_zU_z+m\Omega -\Omega)\omega+(k_zU_z+m\Omega -\Omega)^2+(m-1)\Omega^2-\tilde B_1= 0,
\end{equation}
where
\begin{equation}\label{B1}
\tilde B_1=\left (\omega_\mathrm{Ai}-{{A}\over {\sqrt{4\pi\rho_\mathrm{i}}}}\right )^2 +{{A^2(m-1)}\over {4\pi \rho_\mathrm{i}}}+{{\rho_\mathrm{e}}\over {\rho_\mathrm{i}}} \omega^2_\mathrm{Ae}.
\end{equation}
Here $\omega_\mathrm{Ai}$ is expressed by Eq.~(\ref{omega_A}) with $m>0$. The solution to this equation is
\begin{eqnarray}
\label{eq1}
\omega={{\rho_\mathrm{i}}\over {\rho_\mathrm{i}+\rho_\mathrm{e}}}\left \{\phantom{\sqrt{\frac{\rho_e}{\rho_i}}}\hspace*{-9mm}k_zU_z+m\Omega -\Omega \right. \nonumber \\
 \nonumber \\
 \left.
{}\pm \sqrt{-{{\rho_\mathrm{e}}\over {\rho_\mathrm{i}}}(k_zU_z+m\Omega -\Omega)^2-{{\rho_\mathrm{i}+\rho_\mathrm{e}}\over {\rho_\mathrm{i}}}\left[(m-1)\Omega^2-\tilde B_1 \right ]}\right \}.
\end{eqnarray}

For a negative azimuthal wave mode number, $m<0$, both sides of Eq.~(\ref{disp1}) can be divided by $\sigma^2-\omega^2_\mathrm{Ai}+2(\sigma \Omega+A\omega_\mathrm{Ai}/\!\sqrt{4 \pi \rho_\mathrm{i}})\neq0$, which leads to the second order polynomial
\begin{equation}\label{disp3}
\left (1+{{\rho_\mathrm{e}}\over {\rho_\mathrm{i}}} \right )\omega^2 - 2(k_zU_z-|m|\Omega +\Omega)\omega+(k_zU_z-|m|\Omega +\Omega)^2+(|m|-1)\Omega^2-\tilde B_2= 0,
\end{equation}
where
\begin{equation}\label{B2}
\tilde B_2=\left (\omega_\mathrm{Ai}+{{A}\over {\sqrt{4\pi\rho_\mathrm{i}}}}\right )^2 +{{A^2(|m|-1)}\over {4\pi \rho_\mathrm{i}}}+{{\rho_\mathrm{e}}\over {\rho_\mathrm{i}}} \omega^2_\mathrm{Ae}.
\end{equation}
Here $\omega_\mathrm{Ai}$ is expressed by Eq.~(\ref{omega_A}) with $m<0$. The solution to this equation is
\begin{eqnarray}
\label{eq3}
\omega={{\rho_\mathrm{i}}\over {\rho_\mathrm{i}+\rho_\mathrm{e}}}\left
\{\phantom{\sqrt{\frac{\rho_e}{\rho_i}}}\hspace*{-9mm}k_zU_z-|m|\Omega +\Omega \right. \nonumber \\
 \nonumber \\
 \left.
{}\pm \sqrt{-{{\rho_\mathrm{e}}\over {\rho_\mathrm{i}}}(k_zU_z-|m|\Omega +\Omega)^2-{{\rho_\mathrm{i}+\rho_\mathrm{e}}\over {\rho_\mathrm{i}}}\left[(|m|-1)\Omega^2-\tilde B_2 \right ]}\right \}.
\end{eqnarray}

When the expressions in square roots of Eqs.~(\ref{eq1}) and (\ref{eq3}) are negative, then the system is unstable to KHI. Instability criterion for Kelvin-Helmholtz instability of twisted rotating jets is
\begin{equation}\label{criterion1}
{{\rho_\mathrm{e}}\over {\rho_\mathrm{i}}}(k_zU_z+m\Omega -\Omega)^2+\left (1+{{\rho_\mathrm{e}}\over {\rho_\mathrm{i}}} \right )(m-1)\Omega^2>\left (1+{{\rho_\mathrm{e}}\over {\rho_\mathrm{i}}} \right )\tilde B_1
\end{equation}
for $m>0$ and
\begin{equation}\label{criterion2}
{{\rho_\mathrm{e}}\over {\rho_\mathrm{i}}}(k_zU_z-|m|\Omega +\Omega)^2+\left (1+{{\rho_\mathrm{e}}\over {\rho_\mathrm{i}}} \right )(|m|-1)\Omega^2>\left (1+{{\rho_\mathrm{e}}\over {\rho_\mathrm{i}}} \right )\tilde B_2
\end{equation}
for $m<0$.

In the long wavelength approximation, $k_z a \ll 1$, the real and imaginary parts of unstable mode frequency (for both, positive and negative $m$) are
\begin{eqnarray}
\label{omegas}
\mathrm{Re}(\omega)={{\rho_\mathrm{i}}\over {\rho_\mathrm{i}+\rho_\mathrm{e}}}{{m}\over {|m|}}(|m|-1)\Omega, \nonumber \\
  \nonumber \\
\mathrm{Im}(\omega)=\sqrt{{{\rho_\mathrm{i}}\over {\rho_\mathrm{i}+\rho_\mathrm{e}}}(|m|-1)\left [{{\rho_\mathrm{i}+|m|\rho_\mathrm{e}}\over {\rho_\mathrm{i}+\rho_\mathrm{e}}}\Omega^2-|m|{{A^2}\over {4\pi \rho_\mathrm{i}}}\right ]}.
\end{eqnarray}
The expression of Re($\omega$) shows that the real part of frequency is positive (negative) for positive (negative) azimuthal wave mode number. Therefore, the unstable modes with $m>0$ and $m<0$ propagate in opposite directions, but with the same growth rates. Both real and imaginary parts of frequency tend to zero for $m=1$, i.e., there are no unstable modes with $m=1$ in the long wavelength approximation.

The instability criteria, Eqs.~(\ref{criterion1}) and (\ref{criterion2}), are now replaced by the expression
\begin{equation}\label{criterion3}
{{\rho_\mathrm{i}+|m|\rho_\mathrm{e}}\over {\rho_\mathrm{i}+\rho_\mathrm{e}}}\Omega^2 > |m|{{A^2}\over {4\pi \rho_\mathrm{i}}}.
\end{equation}

In most cases, magnetic jets in the solar atmosphere are denser than surrounding medium. For example, giant tornadoes are associated with prominence legs. Therefore, they are much denser than surrounding coronal plasma. The same situation applies to spicules, which are two orders of magnitude denser than surroundings. Using a dense tube approximation, $\rho_\mathrm{e}/\rho_\mathrm{i}\ll 1$, one can get the frequency, growth rate and instability criterion as
\begin{equation}\label{omegad}
\mathrm{Re}(\omega)={{m}\over {|m|}}(|m|-1)\Omega, \quad \mathrm{Im}(\omega)=\sqrt{(|m|-1)\left [\Omega^2-|m|{{A^2}\over {4\pi \rho_\mathrm{i}}}\right ]},
\end{equation}
\begin{equation}\label{criterion4}
\Omega^2 > |m|{{A^2}\over {4\pi \rho_\mathrm{i}}}.
\end{equation}
The last equation implies that the dense rotating jets are unstable when the kinetic energy of rotation is larger than the magnetic energy of twist multiplied by azimuthal wave mode number. Harmonics with large $m$ are stabilized to the instability. It is clear from physical point of view as the harmonics with larger $m$ have stronger tension force, which acts against the destabilizing role of rotation.

For rarified magnetic jets, $\rho_\mathrm{i}/\rho_\mathrm{e}\ll 1$, the frequency, growth rate and instability criterion are
\begin{equation}
\label{omegar}
\mathrm{Re}(\omega)={{\rho_\mathrm{i}}\over {\rho_\mathrm{e}}}{{m}\over {|m|}}(|m|-1)\Omega, \quad \mathrm{Im}(\omega)=\sqrt{{{\rho_\mathrm{i}}\over {\rho_\mathrm{e}}}|m|(|m|-1)\left [\Omega^2-{{A^2}\over {4\pi \rho_\mathrm{i}}}\right ]}.
\end{equation}
\begin{equation}
\label{criterion4a}
\Omega^2 >{{A^2}\over {4\pi \rho_\mathrm{i}}},
\end{equation}
which means that the jets are unstable when the kinetic energy of the rotation is larger than the magnetic energy of twist for any value of $m$. However, the growth rate is much smaller than in the case of dense jets.

From expressions Eqs.~(\ref{criterion1})--(\ref{criterion2}) one can get the instability criteria for special cases.

\subsection{Helical non-twisted jets: $\Omega \not=0$ and $A=0$}

If the rotation of jet is function of height, then it may twist the axial component of the tube magnetic field and create an azimuthal component. But in the case of uniform rotation the jets are not necessarily twisted, therefore one may consider them as rotating and non-twisted structures.

Then the instability criteria Eqs.~(\ref{criterion1})--(\ref{criterion2}) yield the following expression in the case of negligible longitudinal velocity
\begin{equation}\label{crit4}
{{\rho_\mathrm{i}+|m|\rho_\mathrm{e}}\over {\rho_\mathrm{i}+\rho_\mathrm{e}}}(|m|-1){{a^2 \Omega^2}\over {v^2_{\rm Ai}}} > a^2k^2_z\left (1+{{B^2_{\rm e}}\over {B^2_{z{\rm i}}}}\right ),
\end{equation}
where $v_{\rm Ai}=B_{z{\rm i}}/\sqrt{4\pi\rho_\mathrm{i}}$ is the Alfv\'en speed inside the tube. The instability criterion is different from Eq.~(1) in \citet{Terradas2008}, Eq.~(19) in \citet{Soler2010} and Eq.~(2) in \citet{Antolin2014}. This discrepancy is probably caused by the neglect of centrifugal force in the total pressure boundary condition in those papers.

Using the long wavelength approximation we get
\begin{equation}\label{crit4a}
(|m|-1)\Omega^2 >0,
\end{equation}
which means that non-twisted rotating jets are always unstable to KH instability except harmonics with $|m|=1$.

\subsection{Non-rotating jets in twisted tubes: $A\not=0$ and $\Omega=0$}

The upward motion of a twisted tube may lead to ``vortical illusion.''  This means that the tubes are twisted but not rotating. The same situation can be applied to many other jets in the solar atmosphere. In this case, Eqs.~(\ref{criterion1})--(\ref{criterion2}) yield the following expression
\begin{equation}\label{crit3}
m{{U^2_z}\over {v^2_{\rm Ai}}}>\left (1+{{\rho_\mathrm{i}}\over {\rho_\mathrm{e}}} \right )\left (1+m{{B^2_{\rm e}}\over {B^2_{z{\rm i}}}} \right ),
\end{equation}
which shows that the jet is unstable only for super-Alfv\'enic motions. Note that here we consider the perturbations propagating across the magnetic field, i.e., $mA+k_zB_{z{\rm i}}\approx0$, which have maximal growth rates. However, for a nonmagnetic environment, $B_{\rm e} = 0$, the jet is always unstable for sufficiently large azimuthal wave mode number $m$ \citep{zaqarashvili2010}.

\subsection{Non-rotating jets in non-twisted tubes: $\Omega=0$ and $A=0$}

In this case, Eqs.~(\ref{criterion1})--(\ref{criterion2}) yield the following expression
\begin{equation}\label{crit1}
{{U^2_z}\over {v^2_{\rm Ai}}}>\left (1+{{\rho_\mathrm{i}}\over {\rho_\mathrm{e}}} \right )\left (1+{{B^2_{\rm e}}\over {B^2_{z{\rm i}}}} \right ).
\end{equation}

Therefore, plasma flow along non-twisted magnetic tube is unstable to KHI only for super-Alfv\'enic motions. This is a well-known result \citep{chandrasekhar1961}.

Before we go to the numerical solving the dispersion equation (\ref{dispersion}), it is worth to discuss possible KHI in the presence of MHD waves.

\subsubsection{Torsional Alfv\'en waves}

Torsional Alfv\'en waves have purely azimuthal component of velocity, which has a jump at the tube boundary. Closed boundary conditions lead to torsional oscillations, while very long tubes support propagating modes.

Torsional oscillations in non-twisted homogeneous tubes are expressed as \citep{zaqarashvili2003}:
\begin{equation}
\label{tor11}
u_{\theta}=\alpha v_{\rm Ai}\sin(\omega_n t)\sin(k_n z),
\end{equation}
\begin{equation}
\label{tor12}
b_{\theta}=-\alpha B_{z{\rm i}}\cos(\omega_n t)\cos(k_n z),
\end{equation}
where $\alpha$ is the oscillation amplitude, $\omega_n$ and $k_n$ are the discrete frequency and axial wavenumber of the $n$th mode. At the velocity antinode ($\sin(k_n z)=\pm 1$), where the amplitude of velocity oscillations is maximal, the magnetic field oscillation is zero, therefore Eq.~(\ref{crit4a}) can be easily applied. This implies that the torsional oscillations are always unstable to KHI near velocity antinodes.

On the other hand, propagating torsional Alfv\'en waves are expressed as
\begin{equation}
\label{tor21}
u_{\theta}=\alpha v_{\rm Ai}\cos(\omega t-k_z z),
\end{equation}
\begin{equation}
\label{tor22}
b_{\theta}=-\alpha B_{z{\rm i}}\cos(\omega t-k_z z),
\end{equation}
which means that the velocity and magnetic field oscillations at each $z$ are out of phase, i.e., when the velocity is maximal then the magnetic field is also maximal but with opposite sign. Therefore, in this case Eq.~(\ref{criterion4}) should be applied. Then the azimuthal magnetic field suppresses the KHI. Hence, propagating torsional Alfv\'en waves are stable against the KHI.

\subsubsection{Kink waves}

Kink waves have azimuthal velocity component even without inhomogeneous layer at the surface (thus the resonant absorption is absent), which has a jump at the tube surface, hence it may excite the KHI.

Solutions for azimuthal velocity and magnetic field components of standing and propagating kink waves are described by the similar expressions as Eqs.~(\ref{tor11}) and (\ref{tor12}), and Eqs.~(\ref{tor21}) and (\ref{tor22}), respectively. Therefore, like torsional waves, the standing kink waves are unstable to KHI near the antinodes of azimuthal velocity \citep{Ofman1994,Terradas2008,Soler2010,Antolin2014}. On the other hand, propagating kink waves are stable. However, azimuthal velocity perturbations excited due to the resonant absorption in the inhomogeneous layer may still lead to a KHI for propagating kink waves. This point needs future study.

\section{Numerical solutions to the dispersion equation}

Before starting the numerical task of solving the wave dispersion equation (\ref{dispersion}), we normalize all velocities with respect to the Alfv\'en speed inside the jet, $v_{\rm Ai}$, and all lengths with respect to the tube radius $a$.  In general, as we already said, one assumes that the wave frequency $\omega$ is a complex quantity that implies complex wave phase velocity, $v_{\rm ph} = \omega/k_z$, at real longitudinal wavenumber $k_z$.  The main building quantities in Eq.~(\ref{dispersion}) are the Doppler-shifted frequency, $\sigma$, and the internal Alfv\'en frequency, $\omega_{\rm Ai}$, that can be presented in the forms
\[
    a \sigma = v_{\rm Ai} k_za \left[ \frac{\omega}{k_z v_{\rm Ai}} - M_{\rm A} \left( 1 + m \frac{\varepsilon_2}{k_za} \right) \right]
\]
and
\[
    a \omega_{\rm Ai} = v_{\rm Ai} k_za \left( 1 + m \frac{\varepsilon_1}{k_za} \right),
\]
respectively.  Here $M_{\rm A} = U_z/v_{\rm Ai}$ is the Alfv\'en Mach number, $\varepsilon_1 = B_{\theta }(a)/B_{z{\rm i}}$ is the magnetic field twist parameter, and $\varepsilon_2 = \Omega a/U_z = U_{\theta}(a)/U_z$ is the ``velocity twist parameter.''  There are two input parameters, notably the density contrast $\eta = \rho_{\rm e}/\rho_{\rm i}$ and the ratio of the equilibrium magnetic fields $b = B_{\rm e}/B_{z{\rm i}}$.  Thus, at fixed azimuthal wave mode number $m$, $\eta$, $b$, $\varepsilon_1$, and $\varepsilon_2$, we shall look for complex solutions to the wave dispersion equation (\ref{dispersion}) varying the values of Alfv\'en Mach number, $M_{\rm A}$.  There is also another option, namely to fix $M_{\rm A}$ and search for a pair of $\varepsilon_1$ and $\varepsilon_2$, or practically the ratio $\varepsilon_2/\varepsilon_1$, which would yield unstable solutions.  We will use both approaches.
\begin{figure}[ht]
  \centering
\subfigure{\includegraphics[width = 2.97in]{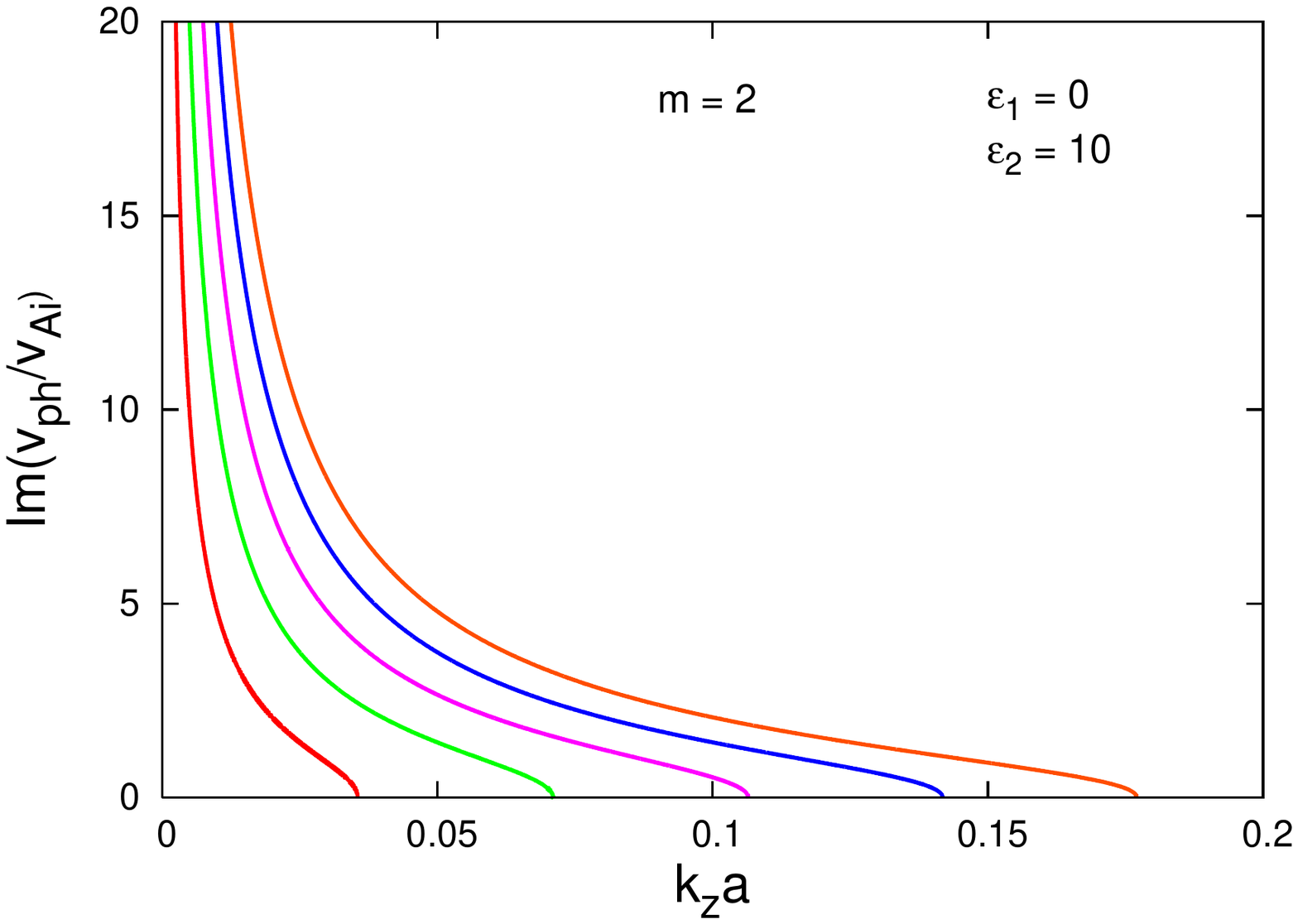}} \hspace{2mm}
\subfigure{\includegraphics[width = 2.97in]{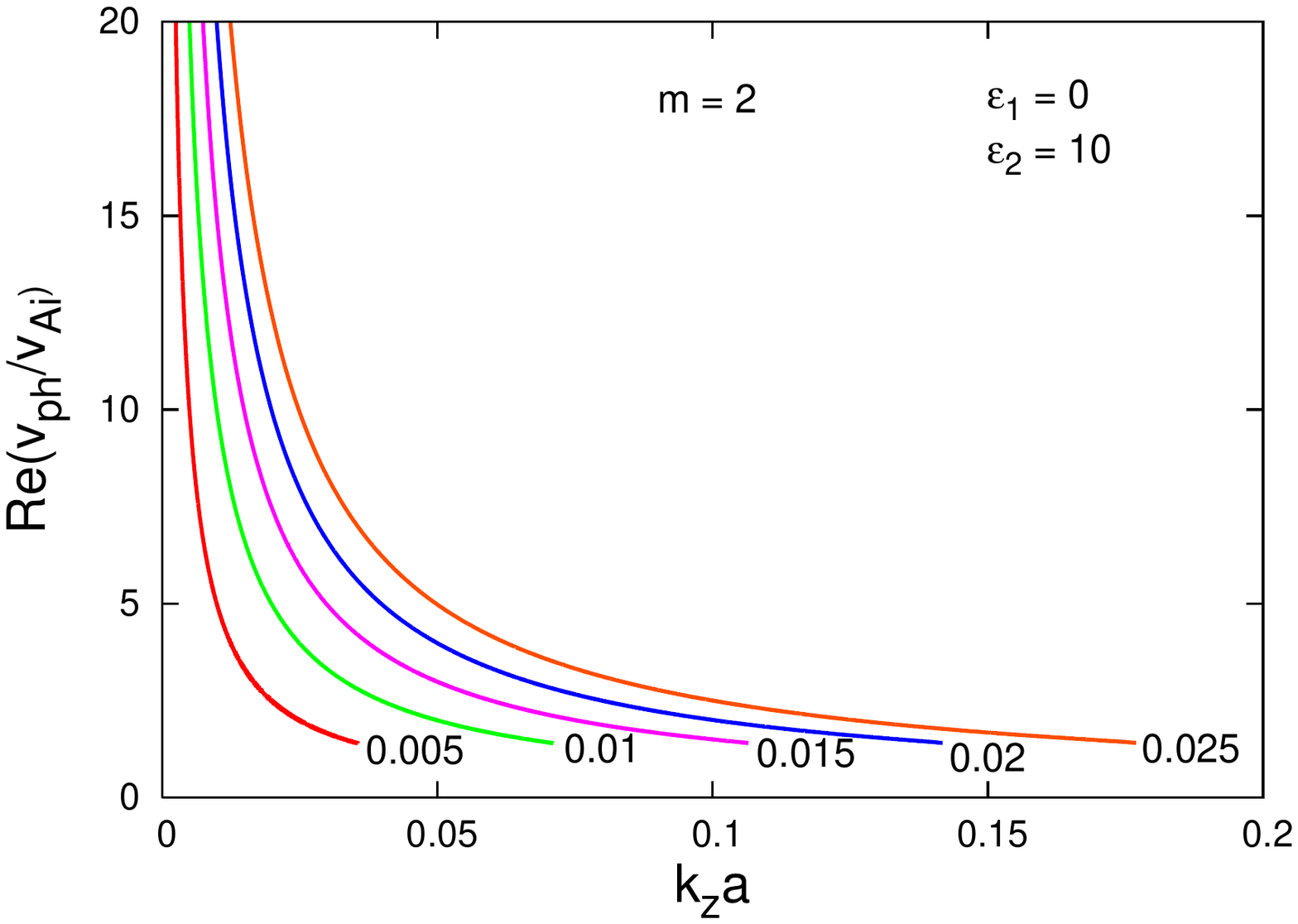}}
  \caption{(\emph{Left panel}) Growth rates of unstable $m = 2$ MHD mode in a helical non-twisted jet at $\rho_{\rm e}/\rho_{\rm i} = 0.01$, $b = 1$, $\varepsilon_2 = U_{\theta}(a)/U_z = 10$, and five various threshold/critical Alfv\'en Mach numbers, $M_{\rm A} = U_z/v_{\rm Ai}$, equal to $0.005$, $0.01$, $0.015$, $0.02$, and $0.025$. (\emph{Right panel}) Dispersion curves of unstable $m = 2$ MHD mode for the aforementioned input parameters.}
  \label{fig:fig1}
\end{figure}

The wave dispersion equation (\ref{dispersion}) and the various instability criteria need a benchmarking.  If we set $\varepsilon_2 = 0$, the numerical solutions
to Eq.~(\ref{dispersion}), for a chosen set of input parameters, strictly coincide with the solutions to, say, Eq.~(13) in \cite{Zhelyazkov2012}.  The instability criterion for KHI in a helical non-twisted jet, $\varepsilon_1 = 0$, for the fluting-like mode $m = 2$ at $\eta = 0.01$, $b = 1$, and $\varepsilon_2 = 10$, according to Eq.~(\ref{crit4}), reads

\begin{equation}
\label{crit4b}
    M_{\rm A} > 0.14073 \, k_z a.
\end{equation}

Computations, performed for five different values of the Alfv\'en Mach number, equal to $0.005$, $0.01$, $0.015$, $0.02$, and $0.025$ yield growth rates and dispersion curves which are shown in Fig.~\ref{fig:fig1}.  The critical dimensionless wavenumbers at which the KHI starts along with the computed from inequality (\ref{crit4b}) critical Alfv\'en Mach numbers are listed in Table~1.
\begin{table}[!h]
\label{tbl:table1}
\caption{Critical wavenumbers for the KHI onset}
\vspace*{3mm}
\begin{tabular}{@{}lcl@{}}
\hline
Line color & Critical $k_z a$ & Computed $M_\mathrm{A}^\mathrm{cr}$ \\
\hline
red        & $0.03552257$ & $0.0049991$ \\
green      & $0.07102204$ & $0.0099949$ \\
purple     & $0.10645792$ & $0.014982$ \\
blue       & $0.14180251$ & $0.019956$ \\
orange     & $0.17701113$ & $0.02491$ \\
\hline
\end{tabular}
\end{table}
As seen, the threshold normalized wavenumbers are in very good agreement with the analytically found instability criterion, Eq.~(\ref{crit4}).

\begin{figure}[ht]
  \centering
\subfigure{\includegraphics[width = 2.97in]{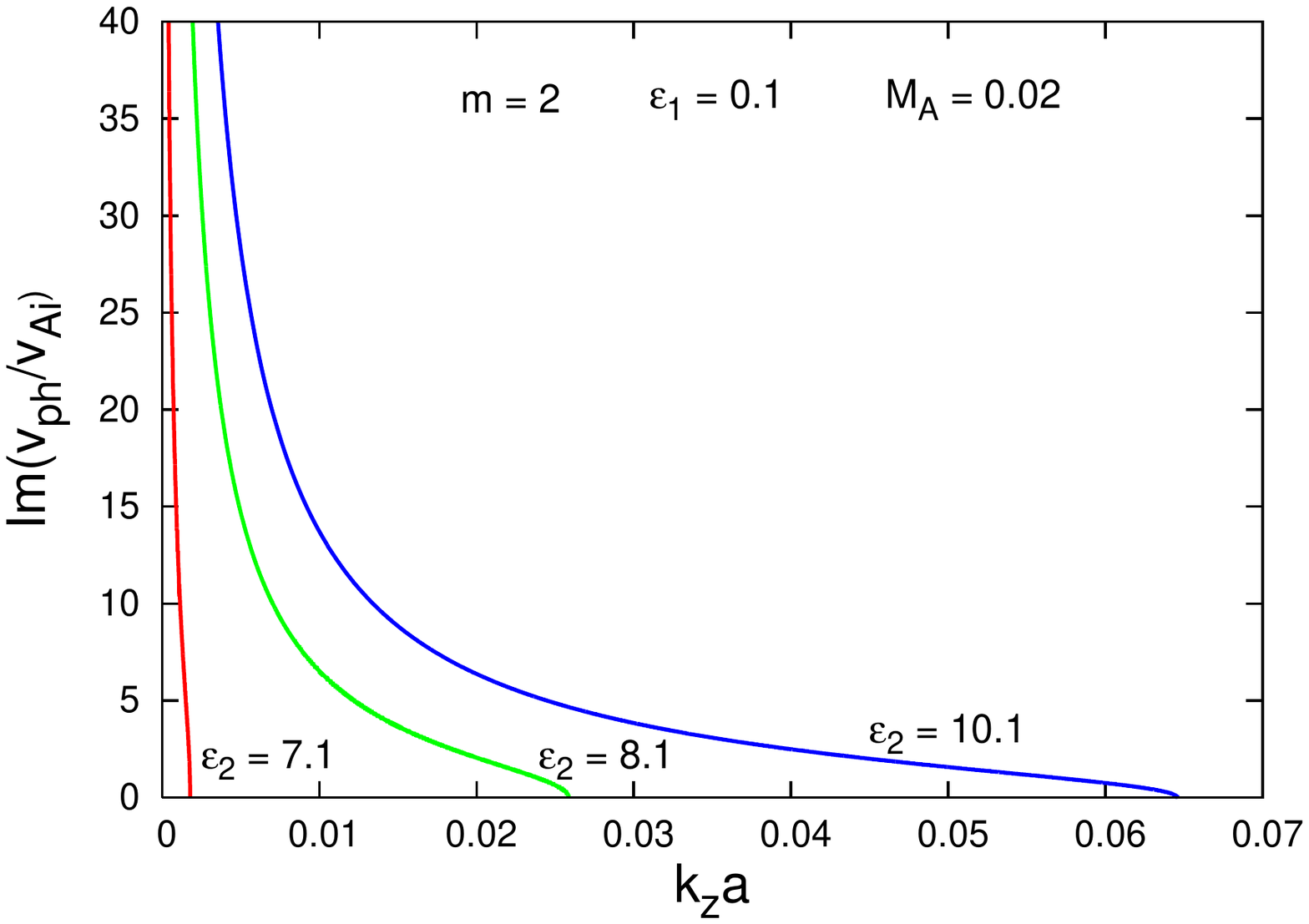}} \hspace{2mm}
\subfigure{\includegraphics[width = 2.97in]{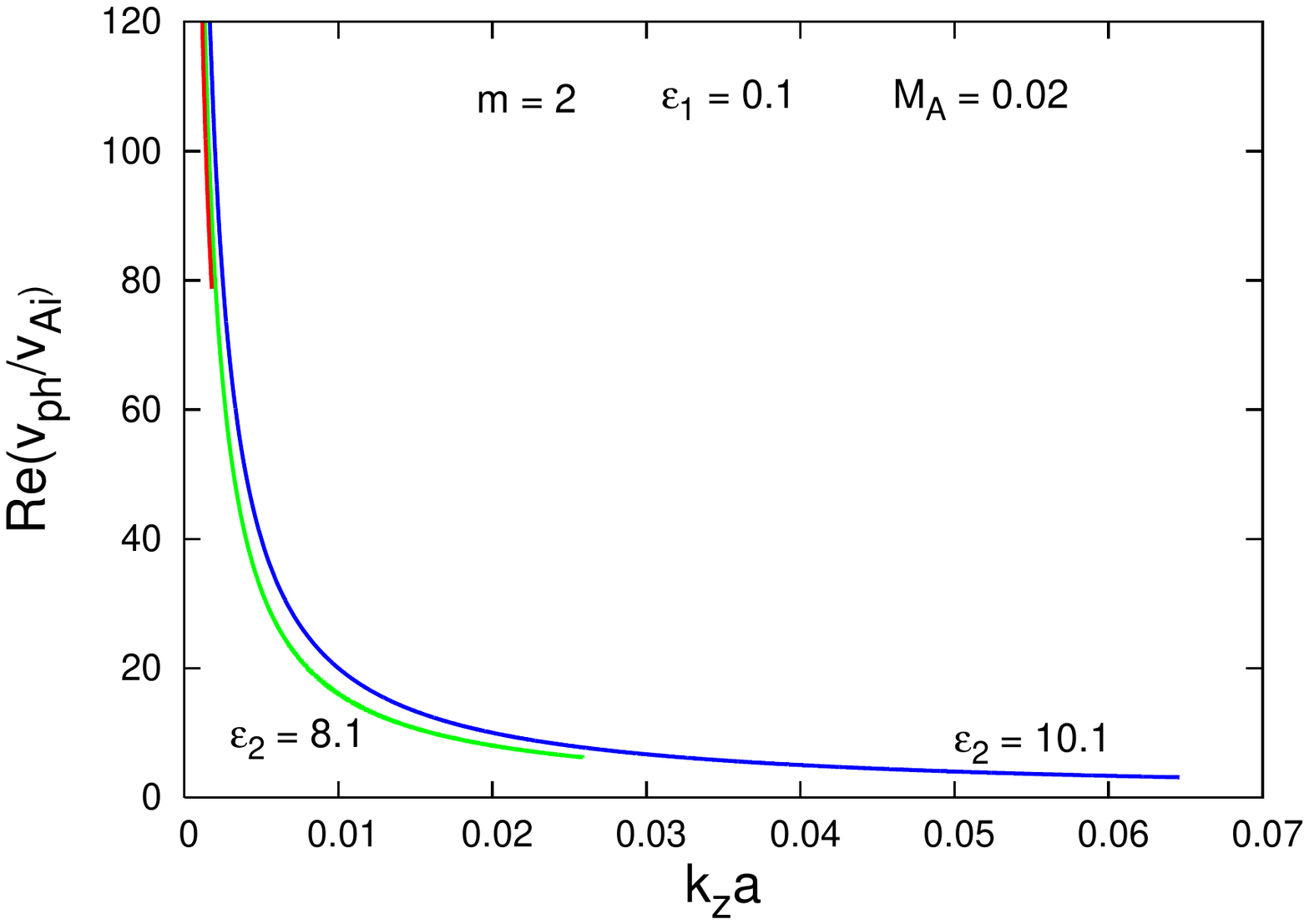}}
  \caption{(\emph{Left panel}) Growth rates of unstable $m = 2$ MHD mode in a helical twisted jet at $\rho_{\rm e}/\rho_{\rm i} = 0.01$, $B_{\rm e}/B_{z{\rm i}} = 1$, $\varepsilon_1 = B_{\theta }(a)/B_{z{\rm i}} = 0.1$, $M_{\rm A} = U_z/v_{\rm Ai} = 0.02$, and three velocity twist parameters, $\varepsilon_2 = U_{\theta}(a)/U_z$, equal to $7.1$, $8.1$, and $10.1$. (\emph{Right panel}) Dispersion curves of unstable $m = 2$ MHD mode for the aforementioned input parameters.}
  \label{fig:fig2}
\end{figure}
\begin{figure}[htpb]
  \centering
\subfigure{\includegraphics[width = 2.97in]{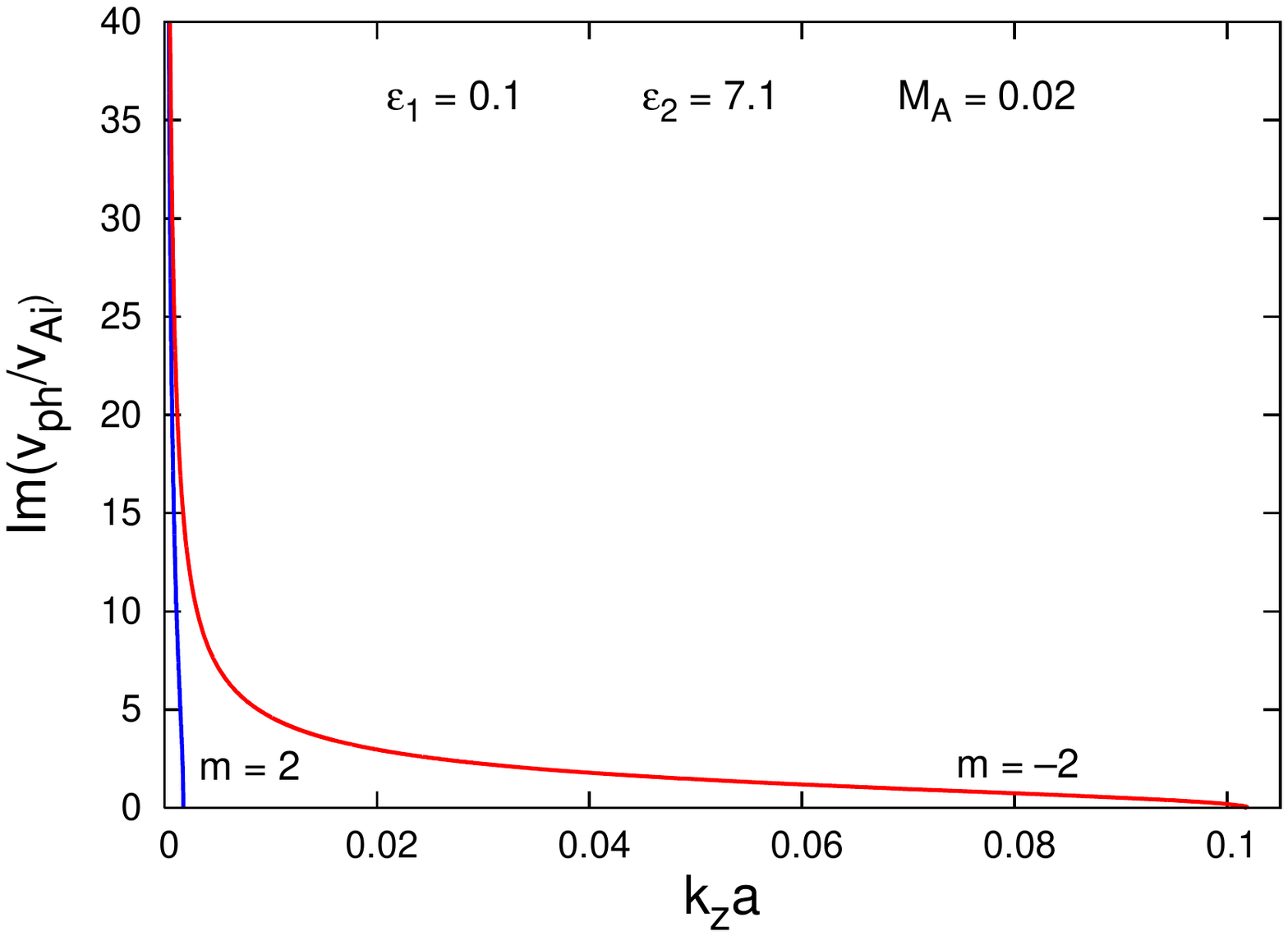}} \hspace{2mm}
\subfigure{\includegraphics[width = 2.97in]{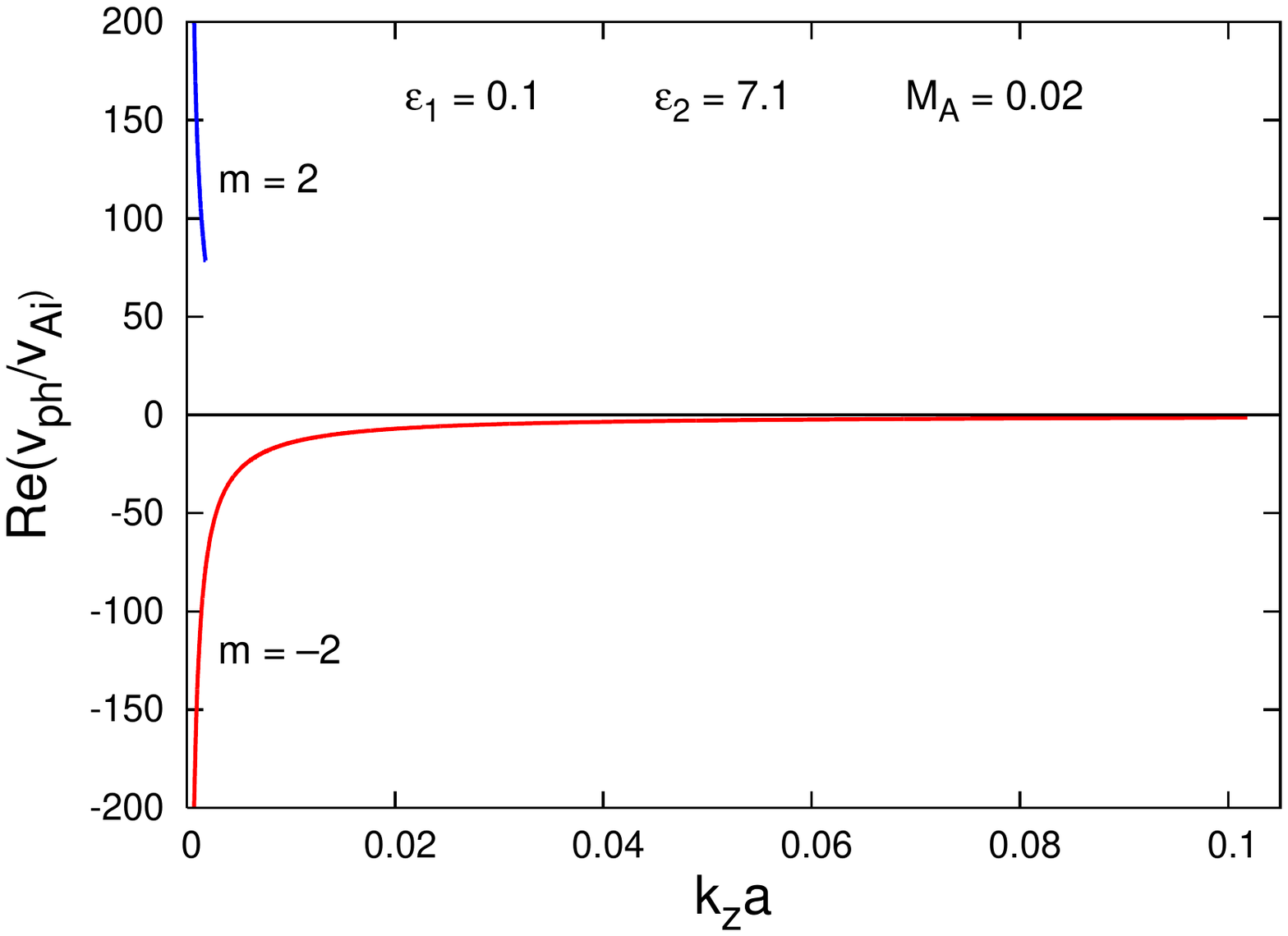}}
  \caption{(\emph{Left panel}) Growth rates of unstable $m = 2$ and $m = -2$ MHD modes in a helical twisted jet at $\rho_{\rm e}/\rho_{\rm i} = 0.01$, $B_{\rm e}/B_{z{\rm i}} = 1$, $\varepsilon_1 = B_{\theta }(a)/B_{z{\rm i}} = 0.1$, $\varepsilon_2 = U_{\theta}(a)/U_z = 7.1$, and $M_{\rm A} = U_z/v_{\rm Ai} = 0.02$. (\emph{Right panel}) Dispersion curves of unstable $m = 2$ and $m = -2$ MHD modes for the aforementioned input parameters.}
  \label{fig:fig3}
\end{figure}
\begin{figure}[htpb]
  \centering
\subfigure{\includegraphics[width = 2.97in]{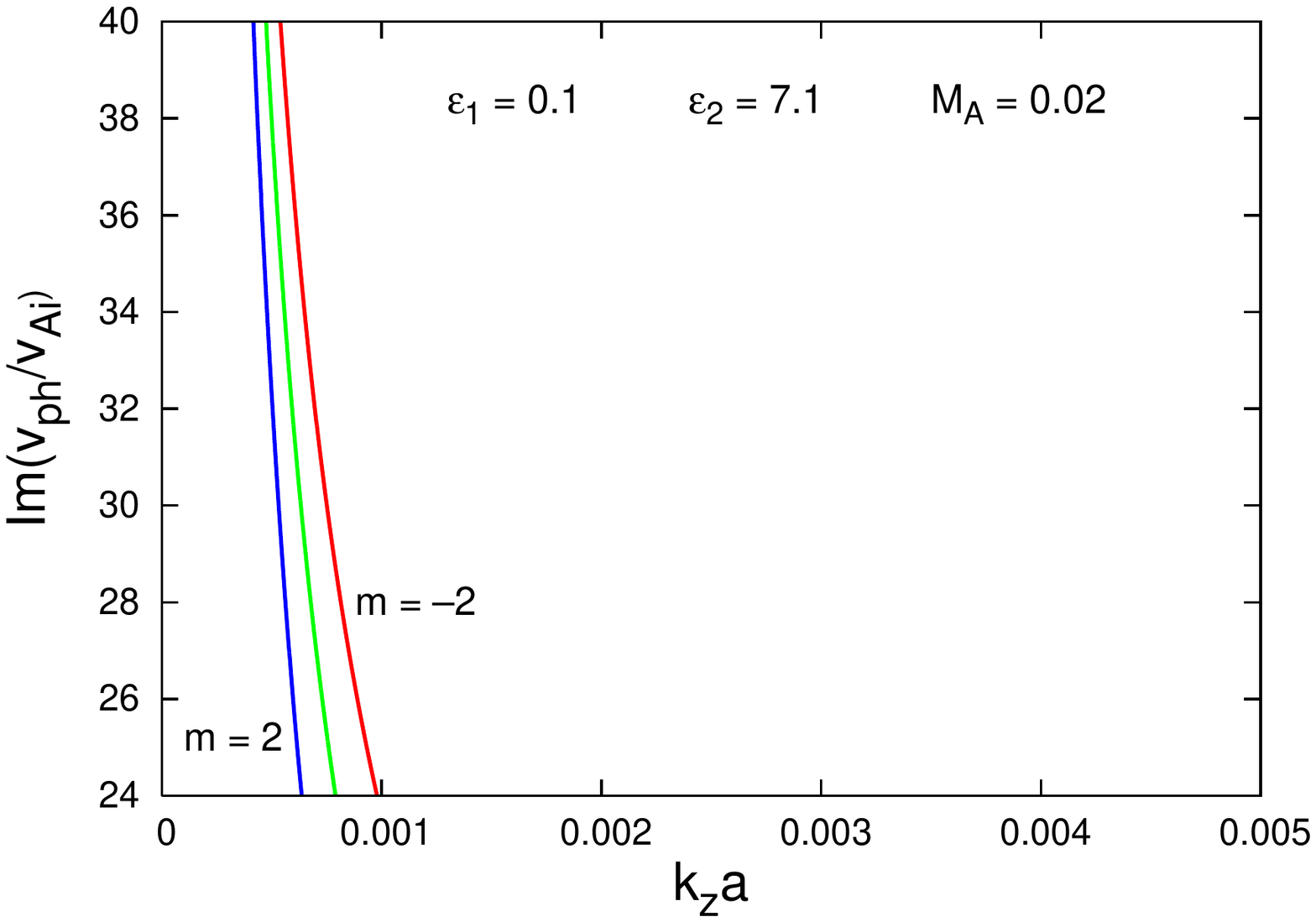}} \hspace{2mm}
\subfigure{\includegraphics[width = 2.97in]{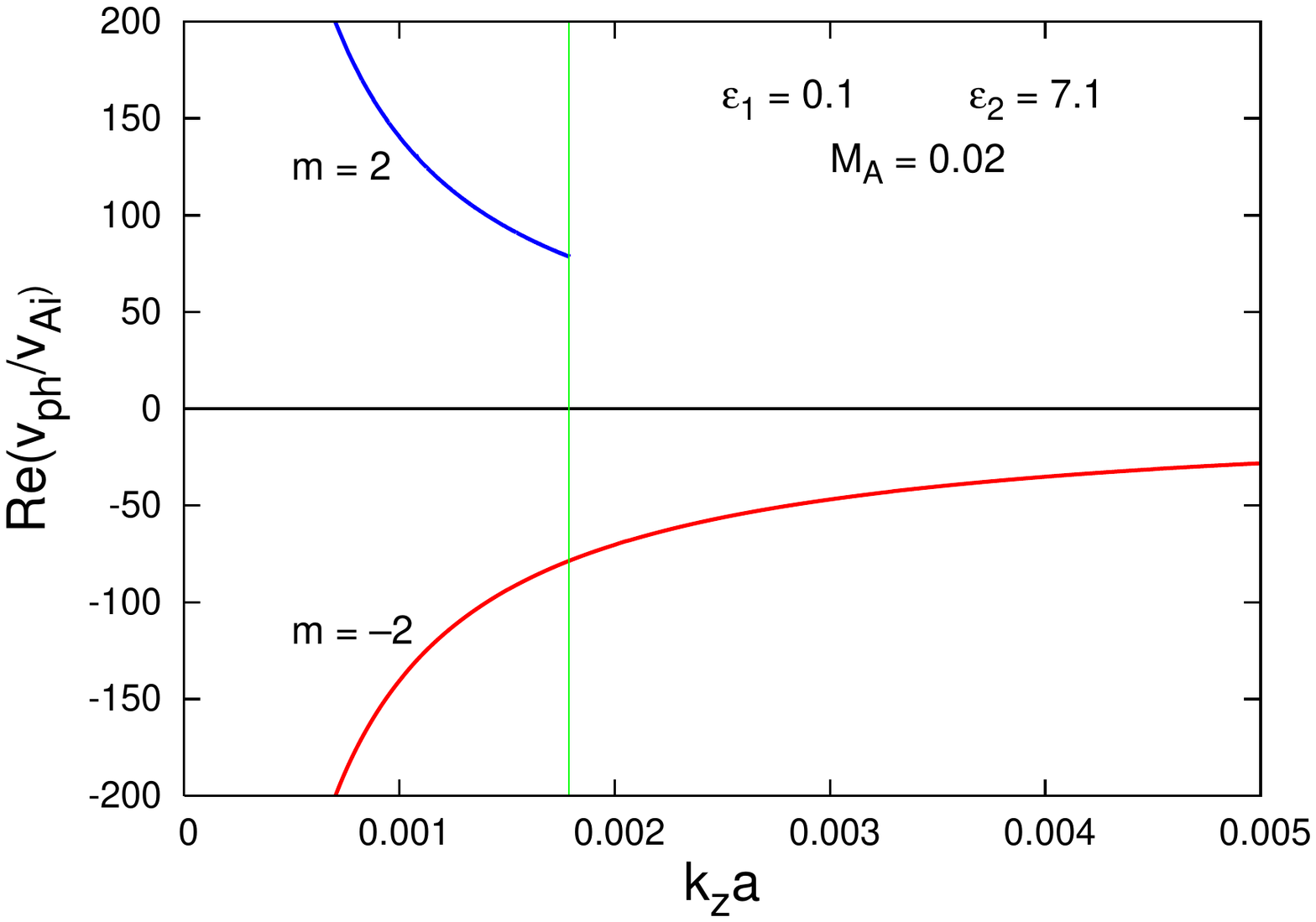}}
  \caption{(\emph{Left panel}) Zoom of the growth rates of unstable $m = 2$ and $m = -2$ MHD modes in a helical twisted jet, plotted in Fig.~\ref{fig:fig3}, for $k_z a \ll 1$.  The green curve is calculated from the expression for Im($\omega$) in Eq.~(\ref{omegas}). (\emph{Right panel}) Zoom of the dispersion curves of unstable $m = 2$ and $m = -2$ MHD modes in the aforementioned narrow $k_z a$-region.}
  \label{fig:fig4}
\end{figure}
The general instability criterion, Eq.~(\ref{criterion3}), in the case of dense jets, $\rho_{\rm e}/\rho_{\rm i} \ll 1$, that is, Eq.~(\ref{criterion4}) can be presented in the form
\begin{equation}
\label{criterion5}
    \frac{\varepsilon_2}{\varepsilon_1} > \frac{\sqrt{|m|}}{M_{\rm A}}.
\end{equation}
If we fix Alfv\'en Mach number to be equal to $0.02$, then one can expect that the $m = 2$ MHD mode would become unstable if $\varepsilon_2/\varepsilon_1 > 70.71$.  Let us set $\varepsilon_1 = 0.1$; then an $\varepsilon_2 = 7.1$ should be sufficient to initiate a KHI in the jet.  Using the same values for the basic input parameters $\eta$ and $b$ as before, we have computed three growth rates and dispersion curves for three different values of $\varepsilon_2$, notably the threshold value of $7.1$ as well as for $8.1$ and $10.1$ -- the results are depicted in Fig.~\ref{fig:fig2}.  The first notable observation is that the instability range of the $m = 2$ mode for $\varepsilon_2 = 7.1$ is extremely narrow.  For the other two values of $\varepsilon_2$ the instability range is much larger, and the values of normalized growth rates and wave phase velocities are relatively lower in a wide $k_z a$-region.

It is interesting to see what happens when the azimuthal wave mode number, $m$, is negative.  In the next Fig.~\ref{fig:fig3} we show the results for the $m = -2$ harmonic at the critical value of $7.1$ of the second twist parameter, compared with those for the $m = 2$ mode.  One can see that, first, the unstable $m = -2$ harmonic is a backward propagating mode, and second, its range of propagation is many times wider than that of the $m = 2$ mode.  We note, however, that in the long wavelength limit, $k_z a \ll 1$, the growth rates of both modes should be, according to Eq.~(\ref{omegas}), the same, and their phase velocities equal, but with opposite signs.  The next Fig.~\ref{fig:fig4} shows a zoom of the very beginning of the $k_za$-axis.  It is easy to observe (see the right panel of Fig.~\ref{fig:fig4}) that the part of the red dispersion curve cut by the green vertical line is a mirror image of the blue dispersion curve of the $m = 2$ mode.  The left panel of the same figure shows that the growth rates of the two modes are not identical, but they are very close to each other.  If we calculate the growth rate from the expression of Im($\omega$) in Eq.~(\ref{omegas}) rearranging it (the expression) in the form
\[
    \frac{\mathrm{Im}(\omega)}{k_z v_{\rm Ai}} = \frac{1}{k_z a}\sqrt{\frac{1}{1 + \eta}\left( \frac{1 + 2\eta}{1 + \eta} \varepsilon_2^2 M_{\rm A}^2 - 2 \varepsilon_1^2 \right)},
\]
the corresponding green curve lies, as is naturally to expect, between the two computed from the wave dispersion relation pieces of growth rate curves.  If one extends the $k_z a$-axis to $0.01$ or $0.02$ the difference between blue and red curves becomes invisible.

\begin{figure}[htpb]
  \centering
\subfigure{\includegraphics[width = 2.97in]{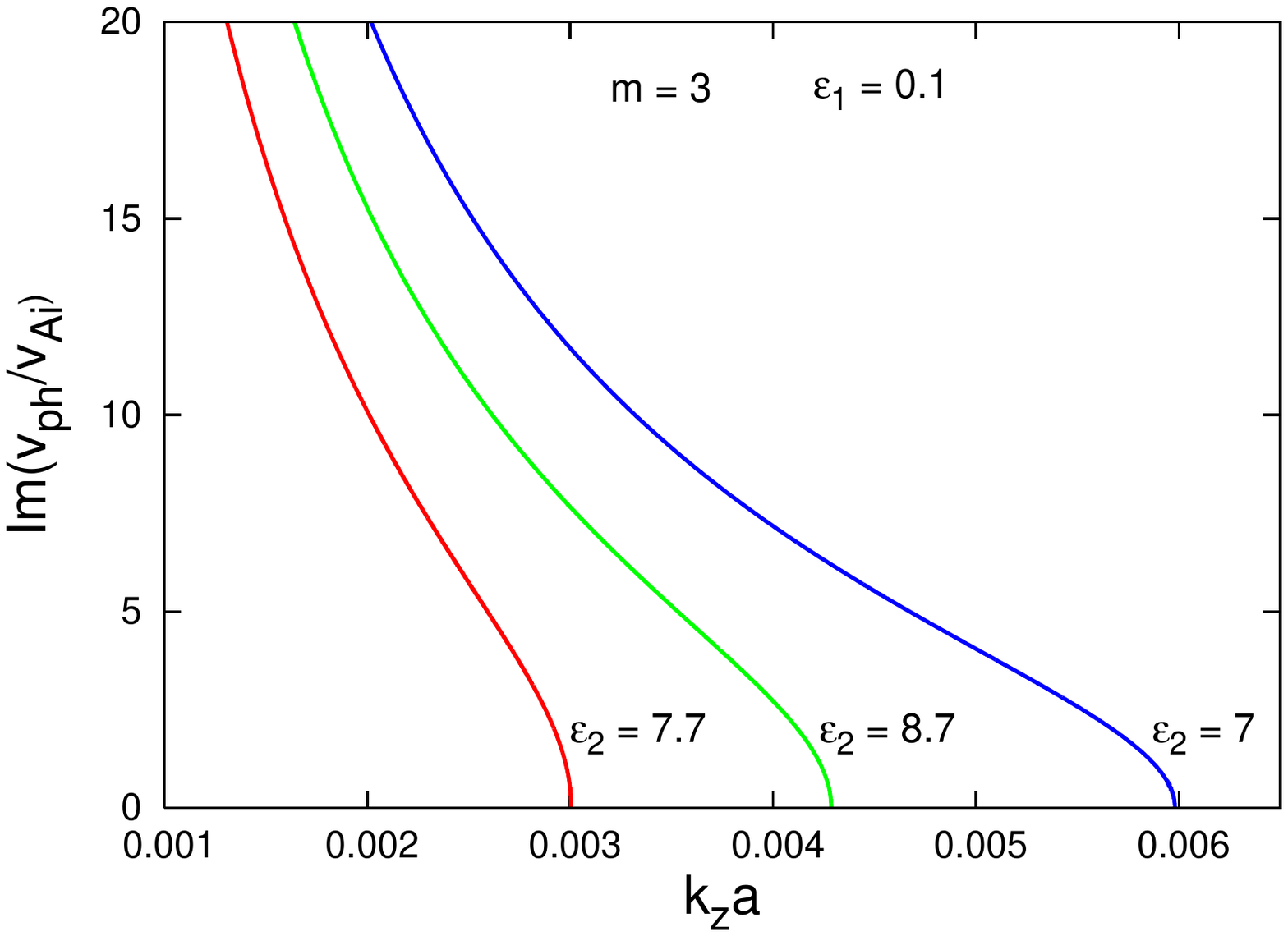}} \hspace{2mm}
\subfigure{\includegraphics[width = 2.97in]{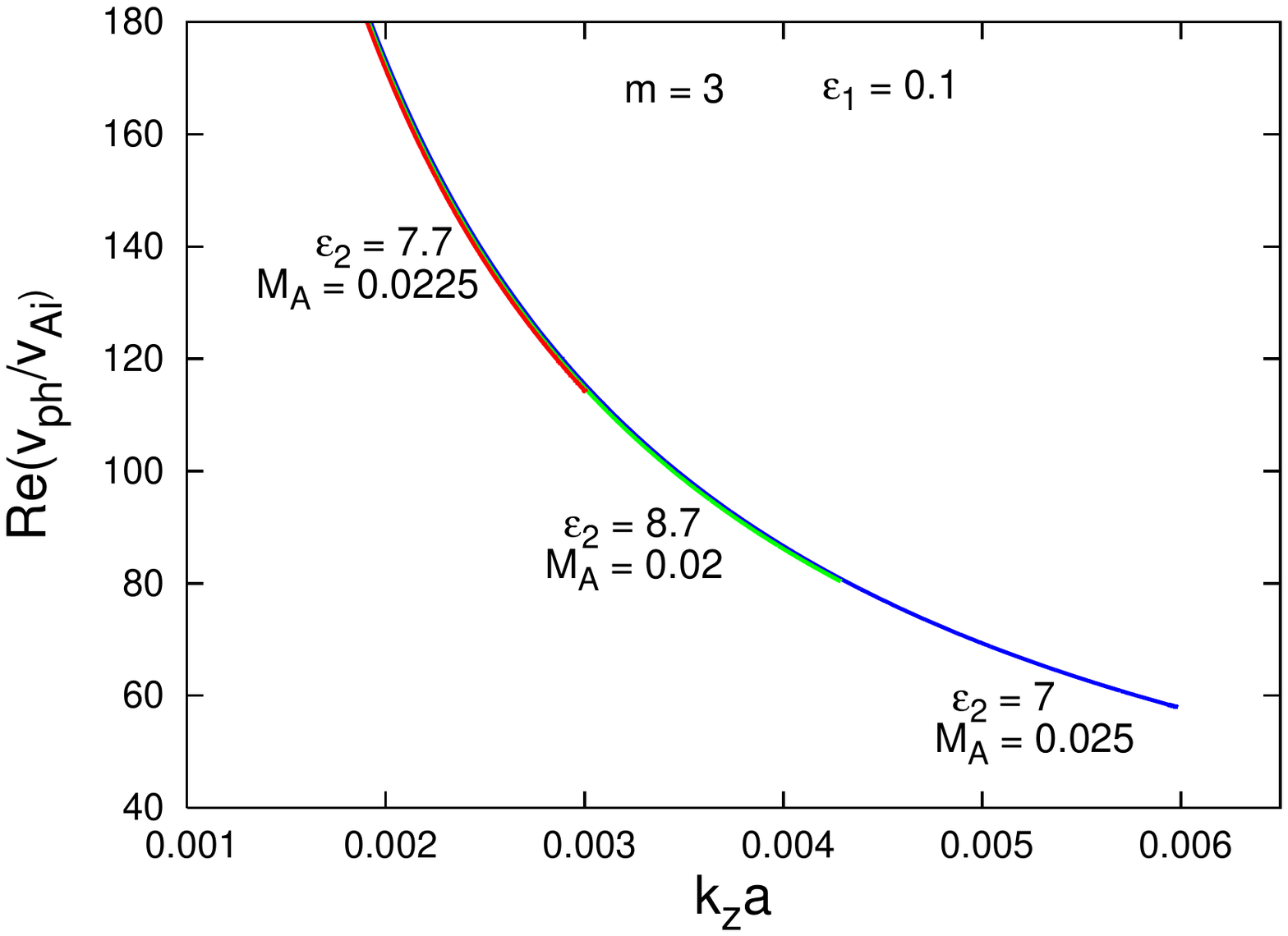}}
  \caption{(\emph{Left panel}) Growth rates of unstable $m = 3$ MHD mode in a helical twisted jet at the same basic parameters as in Fig.~\ref{fig:fig2}, for three Alfv\'en Mach numbers and corresponding to them threshold velocity twist parameters. (\emph{Right panel}) Dispersion curves of unstable $m = 3$ MHD mode for the same input data.}
  \label{fig:fig5}
\end{figure}

One can obtain similar curves for the $m = 3$ harmonic.  Before doing that we will test the general instability criterion, Eq.~(\ref{criterion3}), by choosing three various values of the Alfv\'en Mach number, $M_{\rm A}$, equal to $0.02$, $0.0225$, and $0.025$, respectively.  Assuming the same basic parameters $\eta$ and $b$ as for the fluting-like mode ($m = 2$), we present inequality~(\ref{criterion3}) in the form
\[
    \frac{\varepsilon_2}{\varepsilon_1} > \frac{1.70664}{M_{\rm A}},
\]
which for the fixed $\varepsilon_1 = 0.1$ reduces to
\[
    \varepsilon_2 > \frac{0.170664}{M_{\rm A}}.
\]
This criterion yields the following threshold velocity twist parameters: $\varepsilon_2 > 8.533$ for $M_{\rm A} = 0.02$, $\varepsilon_2 > 7.585$ for $M_{\rm A} = 0.0225$, and $\varepsilon_2 > 6.826$ for $M_{\rm A} = 0.025$.  Note that the criterion for dense jets, Eq.~(\ref{criterion5}), gives a little bit higher values for the threshold ratio $\varepsilon_2/\varepsilon_1$, and subsequently, higher threshold velocity twist parameters.  In particular one obtains $\varepsilon_2/\varepsilon_1 > 86.6$ for $M_{\rm A} = 0.02$, $\varepsilon_2/\varepsilon_1 > 76.98$ for $M_{\rm A} = 0.0225$, and $\varepsilon_2/\varepsilon_1 > 69.28$ for $M_{\rm A} = 0.025$.  We chose for computations the safer values of threshold $\varepsilon_2$ equal correspondingly to $8.7$, $7.7$, and $7$.  The growth rate and wave dispersion curves for these values of $\varepsilon_2$ are plotted in Fig~\ref{fig:fig5}.
\begin{figure}[htpb]
  \centering
\subfigure{\includegraphics[width = 2.97in]{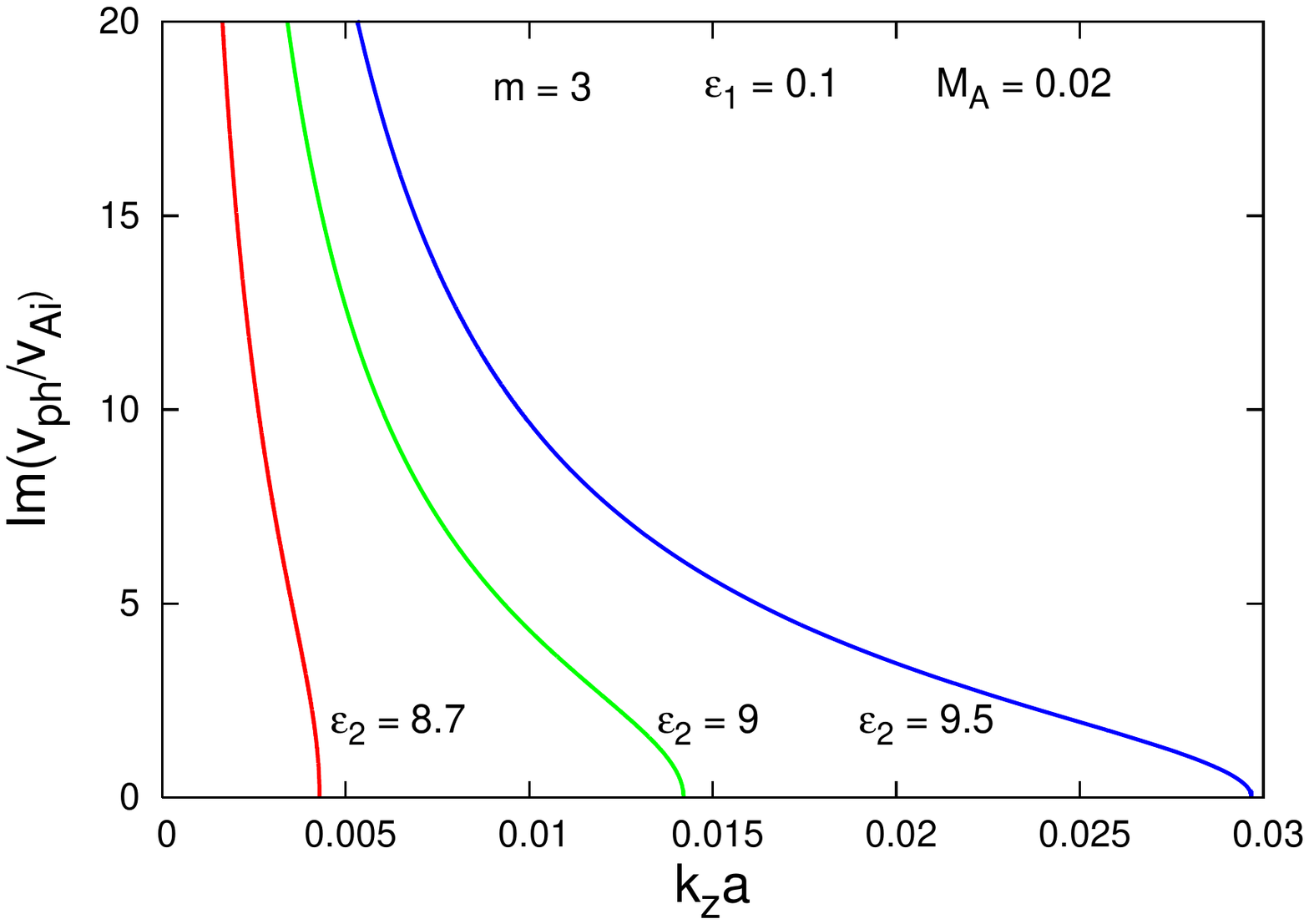}} \hspace{2mm}
\subfigure{\includegraphics[width = 2.97in]{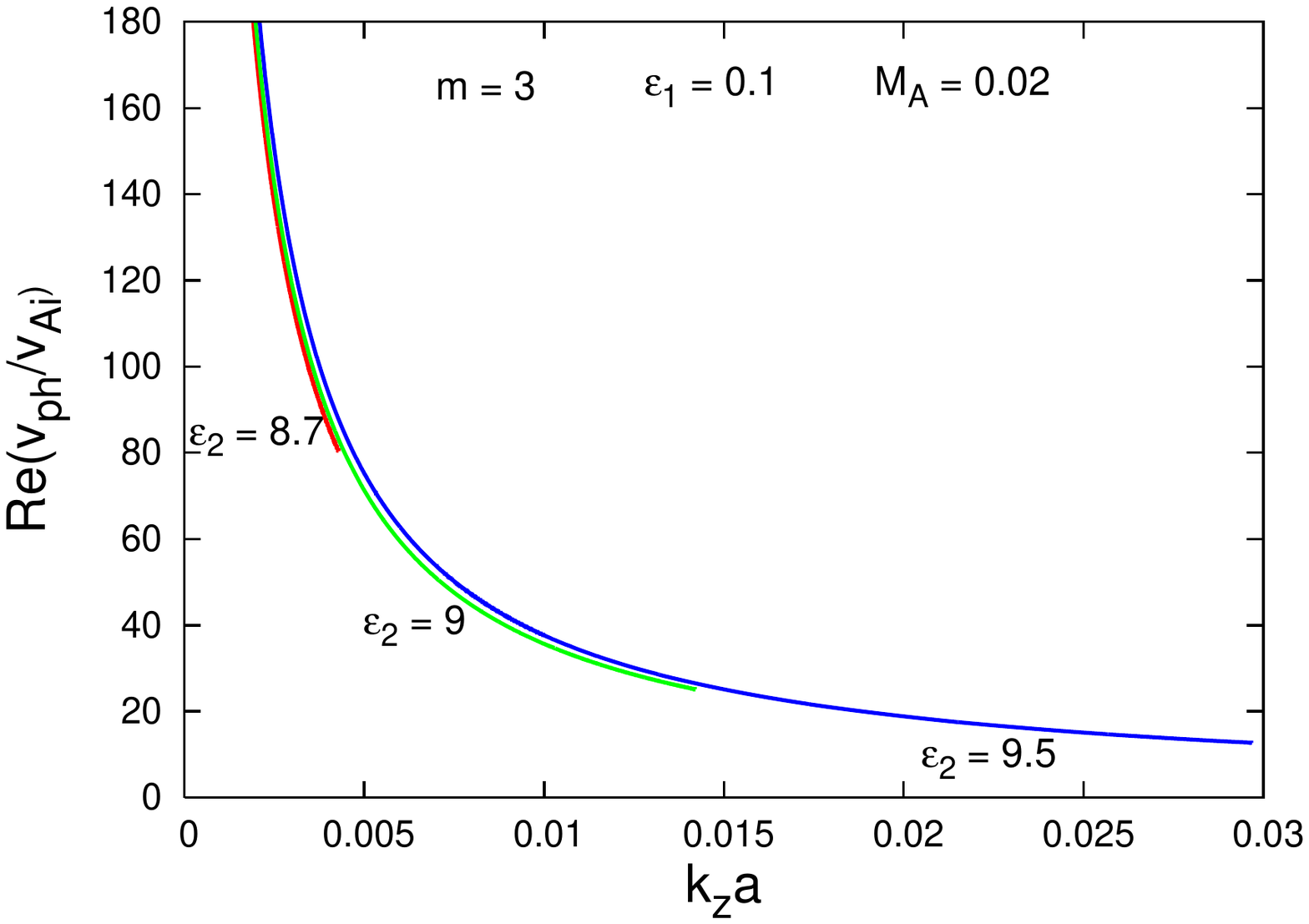}}
  \caption{(\emph{Left panel}) Growth rates of unstable $m = 3$ MHD mode in a helical twisted jet at the same basic parameters as in Fig.~\ref{fig:fig2} for three values of $\varepsilon_2=U_{\theta}(a)/U_z$ equal to $8.7$, $9$, and $9.5$. (\emph{Right panel}) Dispersion curves of unstable $m = 3$ harmonics for the same input data.}
  \label{fig:fig6}
\end{figure}
\begin{figure}[htpb]
  \centering
\subfigure{\includegraphics[width = 2.97in]{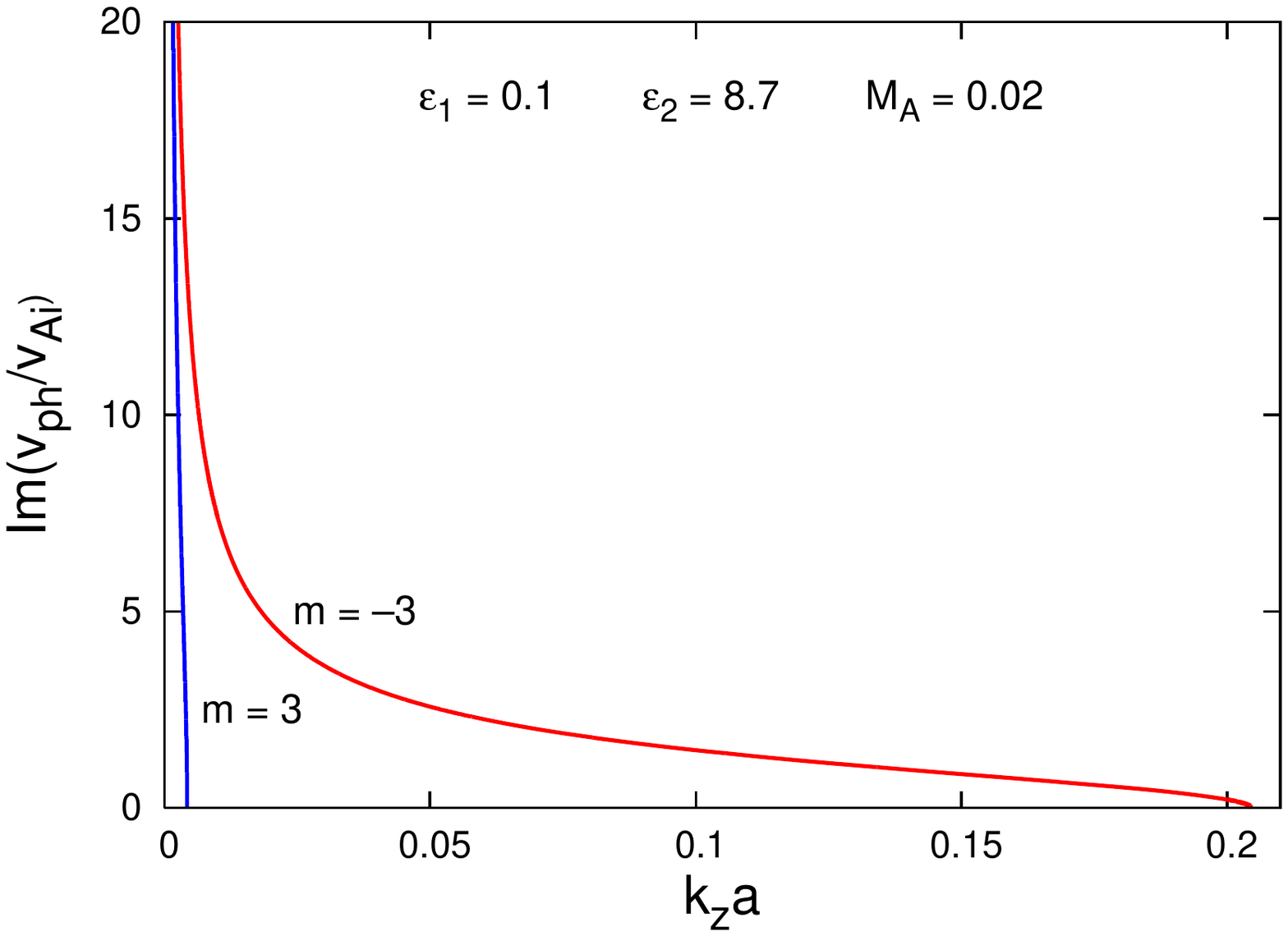}} \hspace{2mm}
\subfigure{\includegraphics[width = 2.97in]{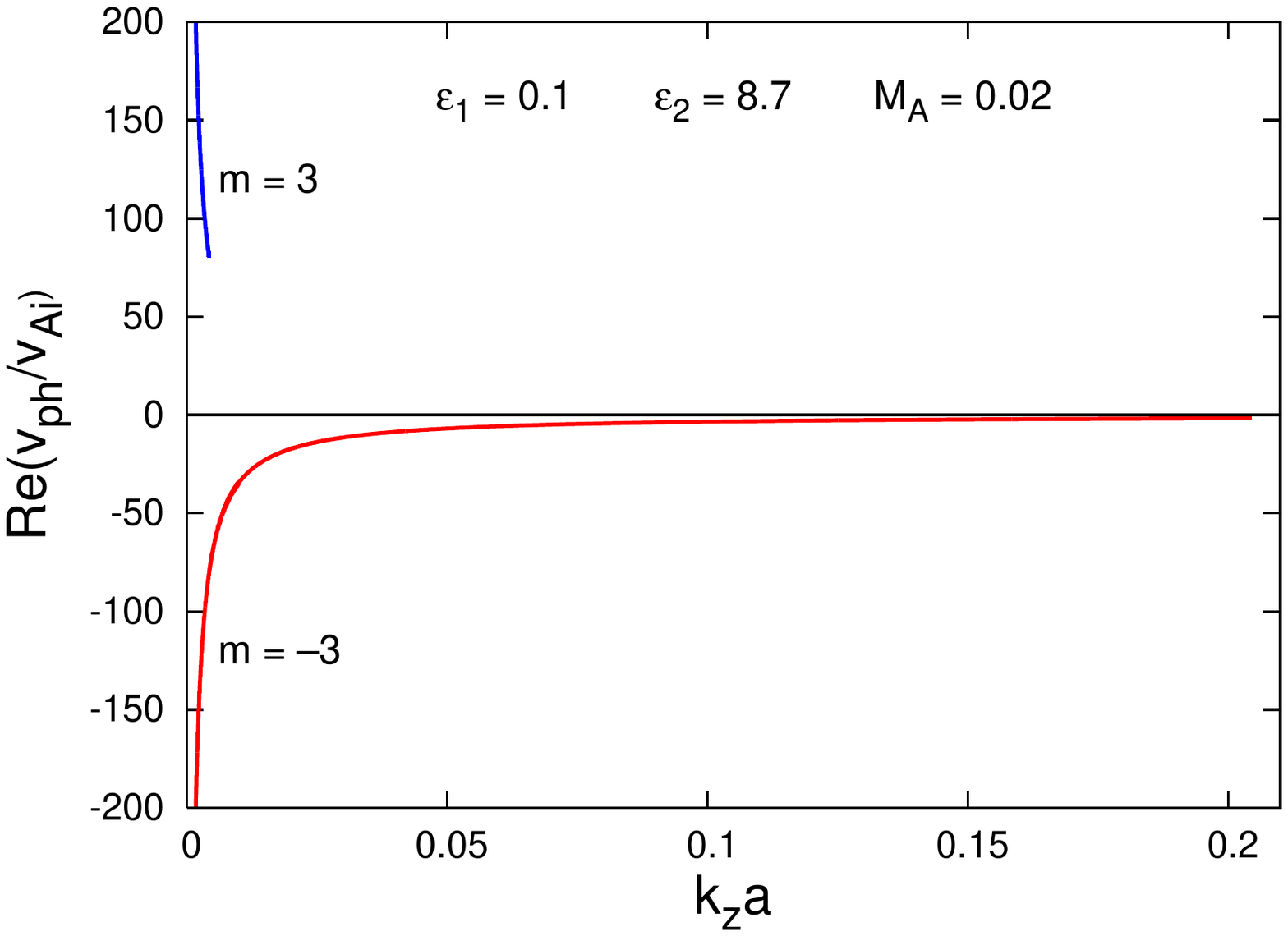}}
  \caption{(\emph{Left panel}) Growth rates of unstable $m = 3$ and $m = -3$ MHD modes in a helical twisted jet at the same basic parameters as in Fig.~\ref{fig:fig6} for $\varepsilon_2 = U_{\theta}(a)/U_z = 8.7$. (\emph{Right panel}) Dispersion curves of unstable $m = 3$ and $m = -3$ MHD modes for the same input data.}
  \label{fig:fig7}
\end{figure}
\begin{figure}[htpb]
  \centering
\subfigure{\includegraphics[width = 2.97in]{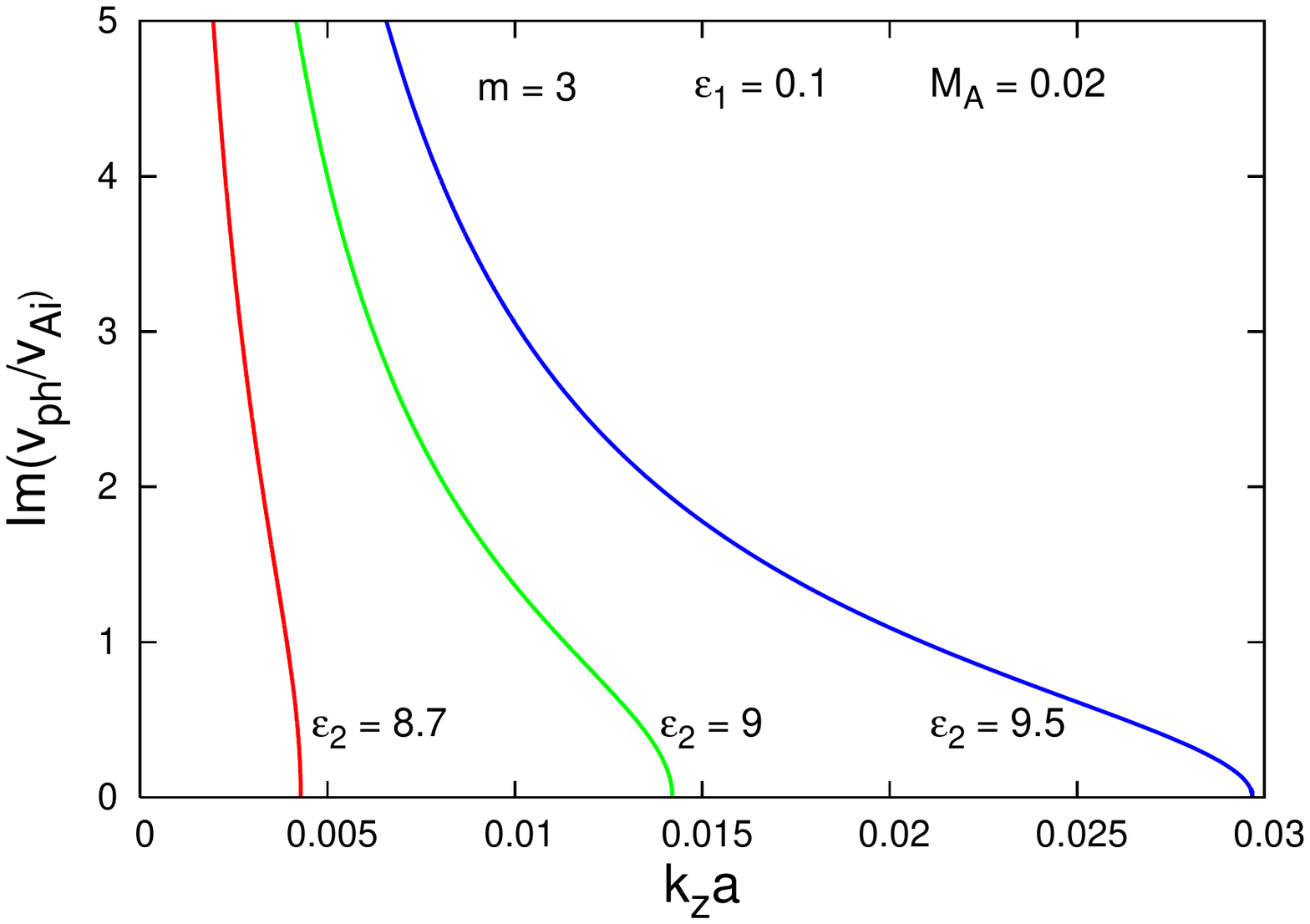}} \hspace{2mm}
\subfigure{\includegraphics[width = 2.97in]{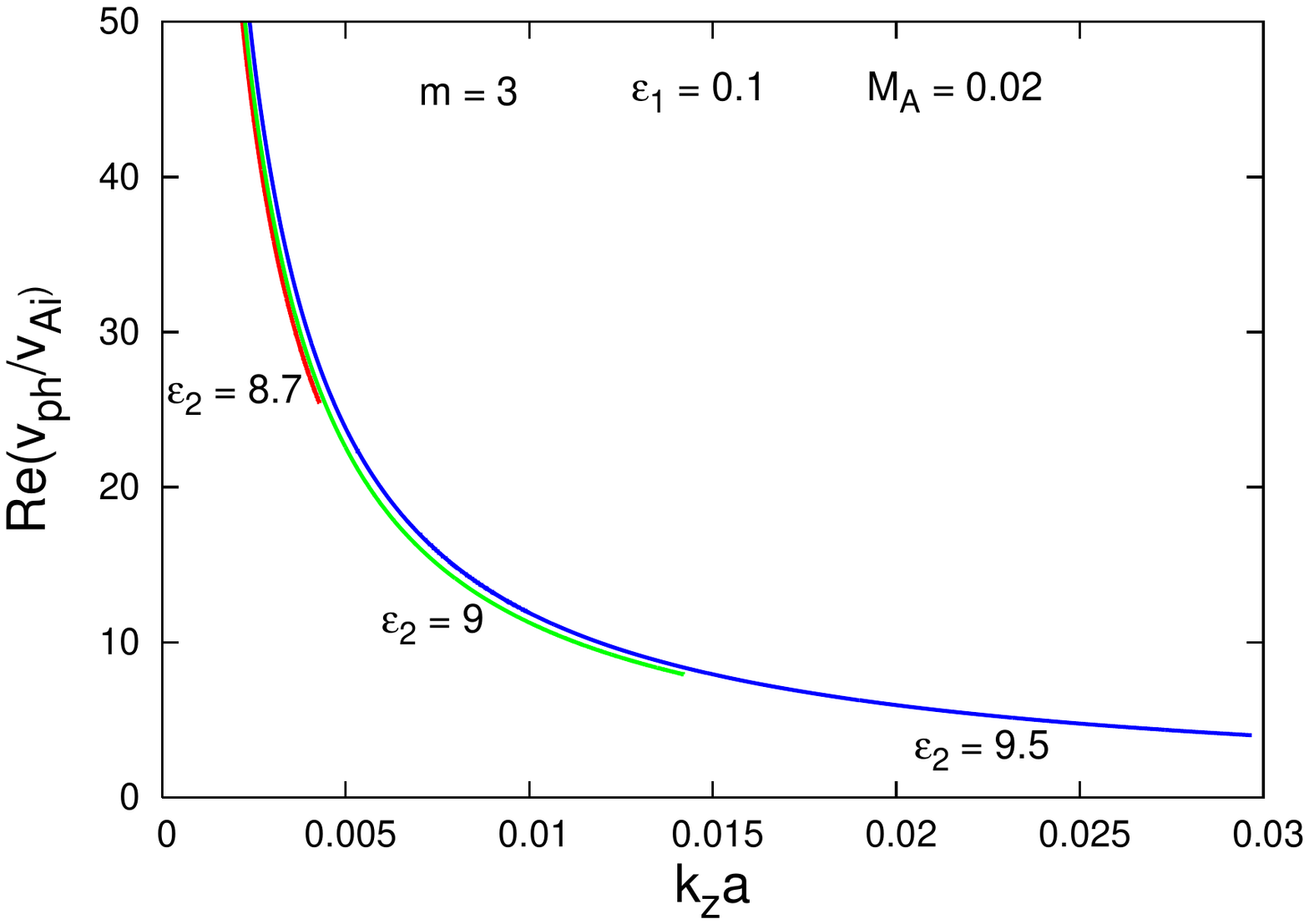}}
  \caption{(\emph{Left panel}) Growth rates of unstable $m = 3$ MHD mode in a helical twisted jet at the same basic parameters as in Fig.~\ref{fig:fig6} for a new definition of Im($v_{\rm ph}$). (\emph{Right panel}) Dispersion curves of unstable $m = 3$ harmonics for the same input data.}
  \label{fig:fig8}
\end{figure}

A set of plots for the $m = 3$ mode, similar to those depicted in Fig.~\ref{fig:fig2}, is shown in Fig.~\ref{fig:fig6}.  It is not surprising the similarity of pictures in these two figures.  The only distinct issue is the circumstance that the instability $k_z a$-range for the $m = 3$ harmonic is wider than that for the fluting-like mode.  Note that while the growth rate curves in both figures are visibly distinguishable, the dispersion curves are very close to each other in their common parts (see the right panels in Figs.~\ref{fig:fig2} and \ref{fig:fig6}).  Moreover, there is an impression that for a fixed value of $M_{\rm A}$ there exists a `universal' dispersion curve along which the dispersion curves for various values of $\varepsilon_2$ lie.

It is not surprising too, that for the $m = -3$ mode we obtain curves similar to those plotted in Fig.~\ref{fig:fig3} as one can see from Fig.~\ref{fig:fig7}.  The common in these two figures is the observation that the normalized wave phase velocities and growth rates in the long wavelength limit, $k_z a \ll 1$, are huge numbers. One can obtain a more reasonable values if we define the actual wave phase velocity as
\[
    v_{\rm ph} = \frac{\omega}{\sqrt{k_z^2 + m^2/a^2}} = \frac{\omega}{k_z \sqrt{1 + m^2/k_z^2a^2}}.
\]
For small $k_z a$, it is the azimuthal propagation (that is, $m$) which defines the wave phase velocity.  With this new definition of $v_{\rm ph}$, the growth rates and dispersion curves plotted in Fig.~\ref{fig:fig6} take the forms shown in Fig.~\ref{fig:fig8}. As seen from that figure, the values of both growth rates and phase velocities are significantly reduced. The same is undoubtedly valid for the $m = 2$ MHD mode, too.
Final conclusion is that the numerical solutions of the dispersion equation (\ref{dispersion}) perfectly agree with analytical instability criteria.

\section{Discussion}

Recent observations with high spatial and temporal resolutions revealed various jets and flows in the solar atmosphere. Some of the observed jets show rotational motion, which may lead to KHI due to jump of azimuthal velocity near boundary. It is obvious that both axial and azimuthal components may take part in the instability depending on the value of velocity speed and the strength of magnetic field in jets. It has been known for a long time that the flow-aligned magnetic field stabilizes sub-Alfv\'enic motions. Therefore, the structure of magnetic field in jets is crucial for possible onset of the instability.

The main limitation of our study is the use of linear incompressible MHD equations in order to derive the dispersion relation governing the dynamics of rotating and twisted tubes. When the perturbations grow further in time, then nonlinear effects as well as secondary instabilities in KH vortices may lead to magnetic reconnection with consecutive heating of plasma. Inclusion of compressibility can have either stabilizing or destabilizing effect, depending on the stationary state of motion \citep{Sen1964}. Therefore, it may change the instability criteria and growth rates. Viscosity may also change the instability criteria and it turns out to have a destabilizing influence when the viscosity coefficient takes different values at the two sides of the discontinuity \citep{Ruderman1996}. All these effects are absent in our simplified model, which only shows the growth of instability in its initial phase in ideal incompressible plasma. Nonlinear and compressible effects may be included in MHD studies of KHI (e.g., \citealp{Miura1984,Ofman2006,Ofman2011,Mostl2013}).

In this study we corrected the boundary condition for Lagrangian total pressure adding the contribution from the centrifugal force, which has been neglected in previous studies (see subsection 2.1). Long wavelength approximation of dispersion equation (\ref{crit4a}) showed that the rotation of non-twisted tubes is always unstable to KHI, but a sufficiently strong twist may stabilize the instability. We found that a twisted rotating rarified jet is unstable to KHI when the kinetic energy of rotation exceeds the magnetic energy of twist (see inequality (\ref{criterion4a})). In the case of dense jet, however, the magnetic energy is multiplied by the azimuthal wave mode number, $m$ (see inequality (\ref{criterion4})). Hence the harmonics with sufficiently high $m$ are always stable. Numerical solutions to the dispersion relation perfectly match with analytical formulas in all cases.

\subsection{Implication for observations}
Instability criteria derived in this paper can be used to estimate the value of azimuthal magnetic field, which may stabilize KHI in various jets with observed rotational motion.

{\it Macrospicules} have typical upward motions of $100$~km\,s$^{-1}$ and rotational speeds of $20$--$50$~km\,s$^{-1}$ \citep{Pike1998}. Then using typical spicule electron density of $10^{11}$~cm$^{-3}$, we estimate the threshold strength of azimuthal magnetic field component from Eq.~(\ref{criterion4}) as $2$--$5$~G (we used most unstable $m = 2$ harmonic for the estimation). Magnetic field strength was measured in spicules by spectropolarimetric observations as ${\sim}10$~G at the height of $2000$~km \citep{Trujillo2005}.  Approximately the same value ($12$--$15$~G) was obtained by \citet{zaqarashvili2007} using observed kink waves in spicules as seismological tools. Recent observations showed even stronger magnetic field reaching $30$~G at $3000$~km height \citep{Orozco2015}. Therefore, the azimuthal magnetic field of $2$--$5$~G, which stabilizes KHI, is below the threshold for the kink instability. Using a magnetic field longitudinal component of $15$--$30$~G leads to an Alfv\'en speed of $100$--$200$~km\,s$^{-1}$. Hence, the observed longitudinal motion of $100$~km\,s$^{-1}$ can be stabilized by a longitudinal magnetic field of $15$--$30$~G. This means that macrospicules will be completely stable to KHI instability for both, axial and azimuthal motions, if the azimuthal component of the magnetic field is ${>}5$~G. However, \citet{Kamio2010} observed the rotational speed of $100$~km\,s$^{-1}$ in a macrospicule, which leads to a threshold value of azimuthal magnetic field of
$10$~G. This is already near the limit of kink instability, which means that the macrospicule could be unstable either to the kink instability if azimuthal magnetic field is stronger than $10$~G or to the KHI if the azimuthal field is less than $10$~G. \citet{Kamio2010} did not observe the kink instability as it might quickly destroy the macrospicule.  Therefore, the azimuthal field should be less than $10$~G. Hence, the macrospicule observed by \citet{Kamio2010} could be unstable to KHI.

{\it Type II spicules} have typical upward motion of $50$--$100$~km\,s$^{-1}$ and torsional motions of $25$--$30$~km\,s$^{-1}$ \citep{De Pontieu2012}. Using the electron number density of $10^{11}$~cm$^{-3}$, one can conclude that the azimuthal component of the magnetic field of ${>}3$~G can stabilize KHI. However, observed strong axial motions, which in some cases could be more than Alfv\'en speed, may lead to KHI in Type II spicules.

{\it X-ray jets} have typical speeds of $200$--$600$~km\,s$^{-1}$, electron number density of 10$^{9}$~cm$^{-3}$ and the temperature of 3$\times$10$^{6}$~K \citep{Shibata1992}. Axial magnetic field of $10$~G leads to an Alfv\'en speed of
$700$~km\,s$^{-1}$, which is higher than the speed of the X-ray jets. This means that the KHI due to the axial motion could be easily stabilized by magnetic field. \citet{Moore2013} observed rotational motions of the order of $60$~km\,s$^{-1}$, which yields the threshold value of azimuthal magnetic field component as $0.6$~G. Therefore, the azimuthal magnetic field of ${>}1$~G stabilizes KHI of rotational motion in X-ray jets observed by \citet{Moore2013}.

{\it EUV jets} have the properties similar to X-ray jets with electron number density of $10^{10}$--$10^{11}$~cm$^{-3}$, but they are rather smaller and short-lived structures. \citet{Zhang2014} observed interesting EUV jet with the initial axial flow of
$250$~km\,s$^{-1}$ and the rotation speed of $120$~km\,s$^{-1}$. Axial magnetic field of $10$~G leads to an Alfv\'en speed of $220$~km\,s$^{-1}$, which is comparable or less than the axial speed of the jets, therefore the axial motion of the jet can be unstable to KHI. For observed rotational motion of $120$~km\,s$^{-1}$, the threshold value of the azimuthal magnetic field component, which may stabilize the KHI of rotational motion, is $4$~G. This implies that the jet should be significantly twisted in order to be stable to KHI. Consequently, both axial and azimuthal velocity components could be unstable in the jet observed by \citet{Zhang2014}.

{\it Giant Tornadoes} are believed to exhibit rotational motions \citep{Wedemeyer2013,Su2014}, however, it was questioned by \citet{Panasenco2014} considering as an illusion effect. There is no common agreement on this topic. \citet{Wedemeyer2013} and \citet{Su2014} estimated the rotation velocity as $20$~km\,s$^{-1}$ and $5$~km\,s$^{-1}$, respectively. For a typical prominence's density of $5 \times 10^{-14}$~g\,cm$^{-3}$, the threshold azimuthal magnetic field can be estimated as $0.3$~G and $1.1$~G for the rotational motions of $5$~km\,s$^{-1}$ and $20$~km\,s$^{-1}$, respectively. Thus the azimuthal component of
${>}1$ G may suppress KHI related to the observed rotation in giant tornadoes. \citet{Mghebrishvili2015} estimated a rising speed of tornado as $1.5$~km\,s$^{-1}$, which is much smaller than the typical Alfv\'en speed in prominences, therefore the axial motion of giant tornadoes probably is also stable to KHI.

%

Besides the jets, KHI can be important in transverse waves and oscillations, which have azimuthal velocity components. Torsional Alfv\'en and kink waves have jumps of azimuthal velocity near tube boundaries and hence can be unstable. The onset of instability depends on spatial structure of azimuthal magnetic field and velocity perturbations. Azimuthal velocity and magnetic field components are in phase difference of $\pi$ for propagating waves, therefore their antinodes coincide with each other (the azimuthal velocity has maximum at the maximum of the azimuthal magnetic field but with opposite sign). Consequently, the azimuthal velocity is stabilized by the azimuthal magnetic field. Therefore, both, propagating Alfv\'en and kink waves are stable to KHI. On the other hand, the azimuthal velocity and magnetic field components are in phase difference of $\pi/2$ for standing waves, therefore the azimuthal magnetic field is zero near at the antinode of velocity. Consequently, the standing kink and torsional Alfv\'en waves are always unstable to KHI near antinodes (see \citealp{Hayvaerts1983}). This phenomenon has been often observed in numerical simulations of standing kink oscillations \citep{Ofman1994,Terradas2008,Antolin2014}.

The jets, which show clear rotational motions, could be twisted in order to suppress KHI. However, if the magnetic field in the jets is not twisted, then the rotation inevitably leads to instability. It is important to estimate the growth time of KHI in the different cases of jets. From Eq.~(\ref{omegas}) one can obtain the growth time for different $m$ harmonics as (here we suppose non-twisted jets)
\begin{equation}
\label{time}
 {{a}\over {U_{\theta}\sqrt{|m|-1}}},
\end{equation}
where $a$ is the radius of jet and $U_{\theta}$ is the azimuthal velocity at the jet boundary. For the macrospicule of \citet{Kamio2010} (radius of $500$~km and rotation speed of $100$~km\,s$^{-1}$), the growth times of $m=2$, $m=3$ and $m=4$  harmonics are $5$, $3.5$, and $2.8$~s, respectively. For the type II spicules of \citet{De Pontieu2012} (radius of $200$~km and rotation speed of $30$~km\,s$^{-1}$) the growth times of $m=2$, $m=3$ and $m=4$  harmonics are $6$, $4.2$, and $3.4$~s, respectively. For the EUV jet of \citet{Zhang2014} (radius of $10\,000$~km and rotation speed of $120$~km\,s$^{-1}$), the growth times of $m=2$, $m=3$ and $m=4$  harmonics are $80$, $56$, and $45$~s, respectively. Hence, the growth time of KHI in spicules and macrospicules is of the order of seconds, while in EUV and X-ray jets is of the order of minutes.

KH vortices excited near the boundary of the jets may lead to plasma turbulence and energy cascade to small scales, which may eventually heat the surrounding plasma. Therefore, the rotating jets could be important for chromospheric and coronal heating. For example, type II spicules show very short live time of the order of $10$--$150$~s and the estimated growth time of KHI is of the order of $1$--$6$~s. Hence the developed turbulence and secondary reconnections due to the KH vortices may lead to the heating of plasma and consequently to the disappearance of the spicules. In this regard, we note recent observations of fast disappearance of RREs and RBEs by \citet{Kuridze2015}, which has been interpreted by the heating of the structures due to the KHI at their boundary during the transverse motion.

\section{Conclusion}

We have studied the KHI of rotating magnetized jets using incompressible magnetohydrodynamic equations. Applying correct boundary condition for Lagrangian total pressure (see subsection 2.1), we obtained dispersion equation, which was solved analytically and numerically. We found that a rotating and twisted tube is unstable to KHI when the kinetic energy of rotation exceeds the magnetic energy of twist. Instability criterion for dense tubes is
$$
U^{2}_{\theta} > |m|{{B^{2}_{\theta}}\over {4\pi \rho_\mathrm{i}}},
$$
where $U_{\theta}$ and $B_{\theta}$ are the rotation speed and azimuthal magnetic field strength at the jet surface, $\rho_\mathrm{i}$ is the jet density and $m$ is the azimuthal wave mode number. On the other hand, the instability criterion for the rarified jet is
$$
U^{2}_{\theta} > {{B^{2}_{\theta}}\over {4\pi \rho_\mathrm{i}}}.
$$
Using observed rotational speed we estimated that the azimuthal magnetic field with strength of $1$--$5$~G can stabilize rotating spicules, macrospicules, X-ray and EUV jets. However, without magnetic twist the rotating jets are always unstable to KHI and the growth time is of the order of several seconds for spicules/macrospicules and of the order of minutes for X-ray and EUV jets. KH vortices may lead to enhanced turbulence and plasma heating due to secondary instabilities, therefore the jets may provide energy source for chromospheric and coronal heating. We also found that standing kink and torsional Alfv\'en waves are always unstable to KHI near their antinodes, while the propagating waves are generally stable. KH vortices may generate enhanced turbulent viscosity, which may contribute to the damping of coronal loop oscillations and need to be studied in the nonlinear regime.

{\bf Acknowledgements} The work of T.V.Z.\ was supported by the Austrian ``Fonds zur F\"{o}rderung der Wissenschaftlichen Forschung'' (FWF) under project P26181-N27, by FP7-PEOPLE-2010-IRSES-269299 project- SOLSPANET and by Shota Rustaveli Foundation grant DI/14/6-310/12.  The work of I.Zh.\ was supported by the Bulgarian Science Fund and the Department of Science \& Technology, Government of India Fund under Indo-Bulgarian bilateral project CSTC/INDIA 01/7, /Int/Bulgaria/P-2/12. The work of L.O.\ was supported by NASA grant NNG11PL10A 670.039.

\appendix

\end{document}